\documentclass[sigplan,screen]{acmart}
\usepackage{amsmath,amsfonts}
\usepackage{physics}
\usepackage{algorithmic}
\usepackage{algorithm}
\usepackage{graphicx}
\usepackage{subcaption}
\usepackage{textcomp}
\usepackage{xcolor}
\usepackage{diagbox}
\usepackage{multirow}
\usepackage{multicol}
\usepackage{makecell}
\usepackage{enumitem}
\usepackage[]{hyperref}

\newcommand{\myFrameworkName}{Fermihedral}
\newcommand{\myFrameworkNameSpace}{Fermihedral }

\def\true{\mathbf{1}}
\def\false{\mathbf{0}}
\DeclareMathOperator{\facomm}{acomm}
\DeclareMathOperator{\weight}{weight}

\copyrightyear{2024}
\acmYear{2024}
\setcopyright{acmlicensed}\acmConference[ASPLOS '24]{29th ACM International Conference on Architectural Support for Programming Languages and Operating Systems, Volume 3}{April 27-May 1, 2024}{La Jolla, CA, USA}
\acmBooktitle{29th ACM International Conference on Architectural Support for Programming Languages and Operating Systems, Volume 3 (ASPLOS '24), April 27-May 1, 2024, La Jolla, CA, USA}
\acmPrice{15.00}
\acmDOI{10.1145/3620666.3651371}
\acmISBN{979-8-4007-0386-7/24/04}

\title{\myFrameworkName: On the Optimal Compilation for Fermion-to-Qubit Encoding}

\author{Yuhao Liu}
\email{liuyuhao@seas.upenn.edu}
\affiliation{%
  \institution{University of Pennsylvania}
  \country{United States}
}

\author{Shize Che}
\email{shizeche@seas.upenn.edu}
\affiliation{%
  \institution{University of Pennsylvania}
  \country{United States}
}

\author{Junyu Zhou}
\email{junyuzh@sas.upenn.edu}
\affiliation{%
  \institution{University of Pennsylvania}
  \country{United States}
}

\author{Yunong Shi}
\email{shiyunon@amazon.com}
\affiliation{%
  \institution{AWS Quantum Technologies}
  \country{United States}
}

\author{Gushu Li}
\email{gushuli@seas.upenn.edu}
\affiliation{%
  \institution{University of Pennsylvania}
  \country{United States}
}

\acmSubmissionID{asplos24fall-p316}



\date{March 2024}

\begin{document}

\begin{abstract}
This paper introduces \myFrameworkName, a compiler framework focusing on discovering the optimal Fermion-to-qubit encoding for targeted Fermionic Hamiltonians. Fermion-to-qubit encoding is a crucial step in harnessing quantum computing for efficient simulation of Fermionic quantum systems. Utilizing Pauli algebra, \myFrameworkNameSpace redefines complex constraints and objectives of Fermion-to-qubit encoding into a Boolean Satisfiability problem which can then be solved with high-performance solvers. To accommodate larger-scale scenarios, this paper proposed two new strategies that yield approximate optimal solutions mitigating the overhead from the exponentially large number of clauses. Evaluation across diverse Fermionic systems highlights the superiority of \myFrameworkName, showcasing substantial reductions in implementation costs, gate counts, and circuit depth in the compiled circuits. Real-system experiments on IonQ's device affirm its effectiveness, notably enhancing simulation accuracy.
\end{abstract}
    
\maketitle 


\section{Introduction}

Simulating Fermionic systems is one crucial application domain of quantum computing. Fermionic systems are composed of Fermions (also known as Fermionic modes), one basic particle type in nature. Notable examples of Fermions include electrons, protons, and neutrons. Many physics models of practical interest are Fermionic systems, such as the molecule electron structure in quantum chemistry~\cite{RevModPhys.92.015003}, the Fermi-Hubbard model~\cite{hubbard} in condensed matter physics and material science, the SYK model~\cite{sykmodel} in quantum field theory. As a fundamentally quantum system, Fermionic systems are hard to simulate on classical computers at a large scale due to their exponential and super-exponential complexity. For example, in 2020, over one million node-hours were allocated to chemistry/material science simulation on the Summit supercomputer~\cite{olcf}, and most of these simulations involve Fermionic systems. 

\begin{figure}[t]
    \centering
    \includegraphics[width=0.8\linewidth]{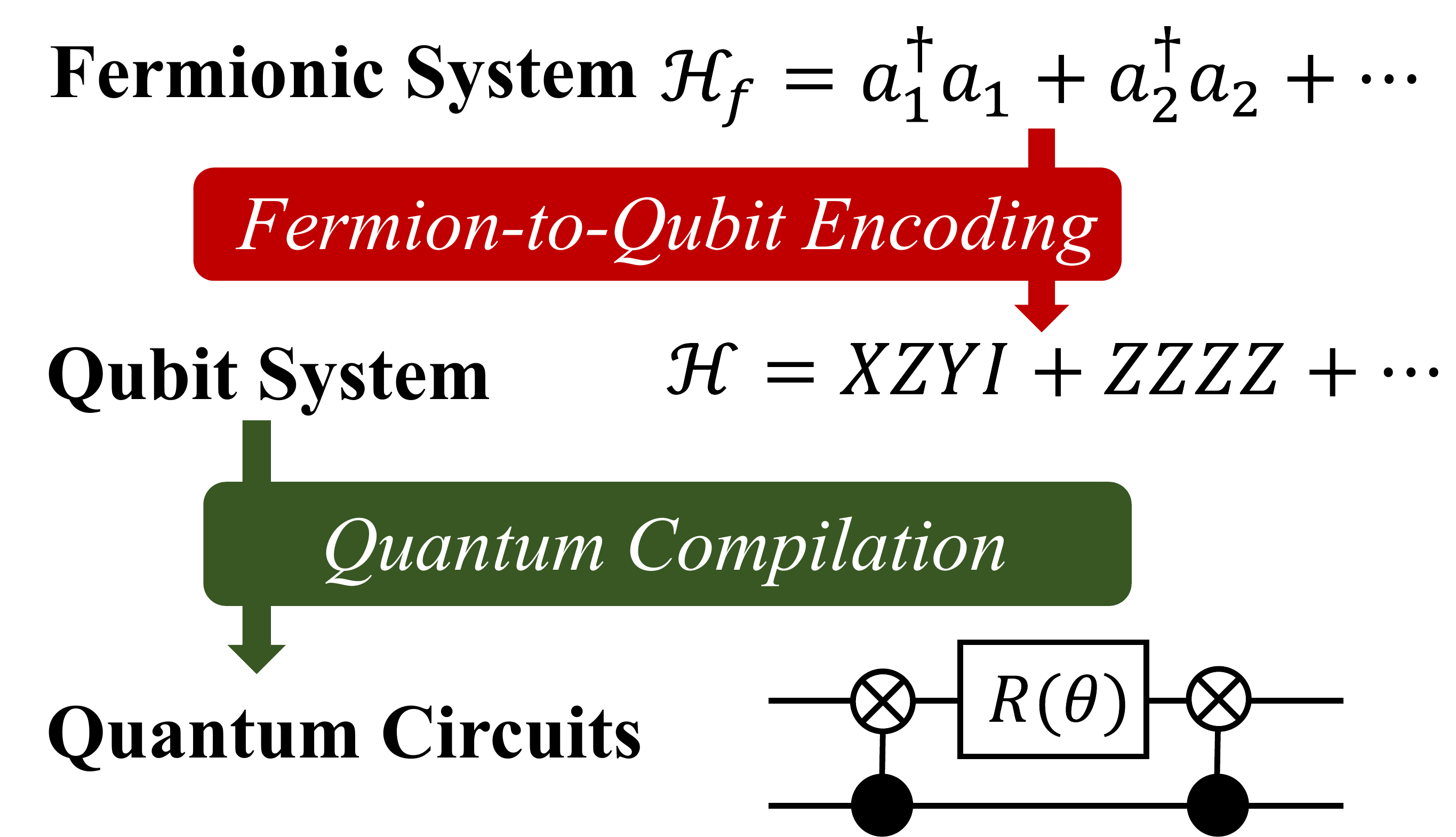}
    \caption{Simulating Fermionic systems with qubit systems}
    \label{fig:fermion-simulate-process}
\end{figure}

Quantum computers are naturally suited to solve such quantum simulation problems. However, encoding a Fermio-nic system onto a quantum computer requires non-trivial efforts. The reason is that most quantum computers are composed of qubits, which satisfy a different statistical property compared with the Fermions in the Fermionic systems. This difference leads to the fact that Fermionic and qubit systems are usually described in two distinct languages. As shown at the top of Figure~\ref{fig:fermion-simulate-process},  the Hamiltonian $\mathit{H}_f$ of a Fermionic system is formulated with an array of creation and annihilation operators $\{a^\dagger_i\}$, $\{a_i\}$ on each Fermionic mode. In the qubit system, however, the Hamiltonian $\mathit{H}$ (in the middle of Figure~\ref{fig:fermion-simulate-process}) is formulated with Pauli string operators (e.g., $XYZI$, $ZZZZ$), which will later be compiled into executable quantum circuits.

A particular transformation called the Fermion-to-qubit encoding is naturally introduced to mitigate the gap between the two disparate languages and encode a Fermionic system onto a quantum computer. This encoding aims to find a set of Pauli strings representing the creation and annihilation operators. These Pauli strings must satisfy a group of constraints to ensure that the unique statistical property of Fermions is preserved in the qubit system. This encoding is not unique, and different encodings will result in very different execution overhead (e.g., gate counts, circuit depth, etc.) on different Fermionic systems.
Overall, it is desirable to have Fermion-to-qubit encodings that can minimize the cost when implementing the quantum circuit to simulate the corresponding Fermionic system.

Finding the optimal Fermion-to-qubit encoding for a targeted Hamiltonian is a highly complicated multi-variable constrained optimization problem.
The constraints on a valid encoding, including the anticommutivity constraints, the algebraic independence constraints, the vacuum state preserving property, and the Hamiltonian implementation cost, are represented in linear algebra and natural number theory. To the best of our knowledge, how to unify and formalize these constraints together is unknown. Existing Fermion-to-qubit encodings are mainly theoretically constructed in a Hamiltonian-independent manner~\cite{jordan_uber_1928}.
Although some encodings~\cite{bravyi_fermionic_2002,miller_bonsai_2022, Jiang_2020} have achieved asymptotically optimal encoding, it is still far from the optimal actual cost because the Hamiltonian of Fermionic systems from various domains can be very different.

\begin{figure}[t]
    \centering
    \includegraphics[width=0.9\linewidth]{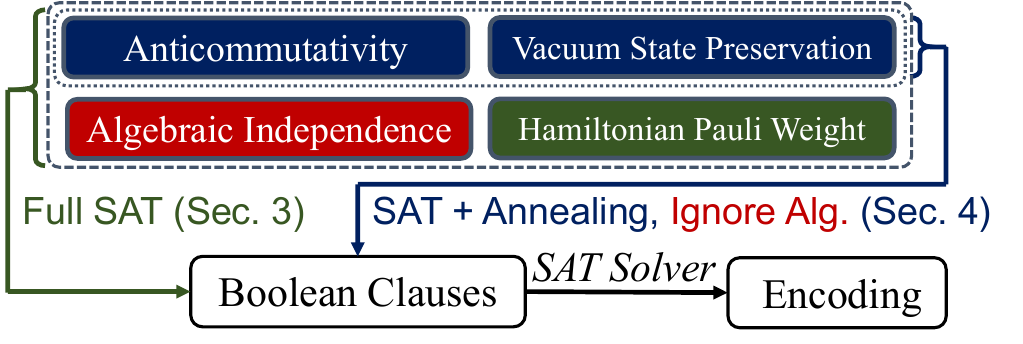}
    \caption{Overview of \myFrameworkNameSpace framework}
    \label{fig:method-process}
\end{figure}

In this paper, we overcome this challenge and propose \myFrameworkName, a compiler framework to find the actual \textbf{optimal} Fermion-to-qubit encoding for a targeted Fermionic Hamiltonian.
The overview of \myFrameworkNameSpace is shown in Figure~\ref{fig:method-process}.
\textbf{First}, by leveraging the Pauli algebra, we can simplify and convert all the constraints represented in linear algebra and natural number theory into Boolean variables and expressions with carefully designed encoding.
Then, the optimal Fermion-to-qubit encoding compilation can be formalized into a Boolean Satisfiability (SAT) problem and solved with existing high-performance SAT solvers.
\textbf{Second}, we identify the immediate bottleneck in our SAT formulation, the exponentially large number of clauses.
We find two causes for this problem and propose corresponding techniques, \textit{ignoring algebraic independence} and \textit{simulated annealing on Hamiltonian-independent optimal encoding}.
These two techniques can accommodate larger-scale cases by providing approximate optimal solutions with negligible failing probability.

We perform a comprehensive evaluation of \myFrameworkNameSpace on various Fermionic systems.
The results show that our SAT-generated optimal Fermion-to-qubit encoding can outperform existing asymptotical optimal encodings~\cite{bravyi_fermionic_2002} and widely adopted encoding~\cite{jordan_uber_1928} with $10\%\sim60\%$ lower Hamiltonian implementation cost, $15\%\sim35\%$ lower gate count and $15\%\sim60\%$ lower circuit depth in the final compiled circuits, as well as more precise Fermionic system simulation results on noisy classical simulators. In particular, we perform real-system experiments showing that our optimal Fermion-to-qubit encoding can significantly increase the simulation accuracy on IonQ's ion trap device.

The major contributions of this paper can be summarized in the following:
\begin{enumerate}
    \item We propose \myFrameworkName, a compilation framework to find the actual \textit{optimal} Fermion-to-qubit encoding for a targeted Fermionic system Hamiltonian.
    \item By leveraging the Pauli algebra, we formulate all the required constraints and implementation costs into Boolean variables and expressions so that the encoding can be solved with an SAT solver.
    \item We propose two techniques to remove unnecessary clauses in our SAT formulation. This allows us to find approximate optimal solutions with negligible failing probability for larger-size cases.
    \item Experimental results show that \myFrameworkNameSpace can outperform current asymptotic optimal encodings on both Hamiltonian-dependent and independent Pauli weight, embodied by better simulation accuracy in noisy simulation and real-system study.
\end{enumerate}


\section{Background}
\label{sec:background}
This section briefly introduces the essential concepts and their properties to help understand this paper. We start with the Pauli strings, the key components in quantum simulation, followed by the introduction to the Fermionic quantum systems. We do not cover basic quantum computing concepts (e.g., qubit, gate, linear operator, circuit) and we recommend~\cite{quantum-computation} for more details.

\subsection{Pauli String}

In quantum simulation, Hamiltonians are usually represented by their decomposition into the sum of Pauli strings. A $N$-length Pauli string for an $N$-qubit system is defined as the tensor product of Pauli operators: $P = \sigma_{N} \otimes \sigma_{N-1} \otimes \cdots \otimes \sigma_{1}$, where $\sigma_i \in \{I, X, Y, Z\}$. Each Pauli operator $\sigma_i$ operates on the qubit $i$ independently.
The $X$, $Y$, and $Z$ are three Pauli operators and $I$ is the identity operator:
$$X=\begin{pmatrix}
    0 & 1 \\
    1 & 0
\end{pmatrix}, 
Y=\begin{pmatrix}
    0 & -i \\
    i & 0
\end{pmatrix},
Z=\begin{pmatrix}
    1 & 0\\
    0 & -1
\end{pmatrix},
I=\begin{pmatrix}
    1 & 0\\
    0 & 1
\end{pmatrix}$$

\subsubsection{Completeness}
\label{sec:pauli-string-orthogonal}
All the length $N$ Pauli strings formulate an orthonormal basis for all the Hamiltonians of $N$ qubits. Formally, for all $N$-qubit Hamiltonian $\mathcal{H}$, there is a unique linear decomposition over all length $N$ Pauli strings:
$$\mathcal{H} = \sum_i w_iP_i, \text{where } w_i \in \mathbb{R}, P_i \in \{I,X,Y,Z\}^{\otimes N}$$

\subsubsection{Pauli String to Circuit}
\label{sec:pauli-string-to-circuit}

The goal of quantum simulation is to implement the operator $exp({i\mathcal{H}t})$. It is usually hard to directly implement this many-qubit unitary operator in the circuit, and we need to compile the $exp({i\mathcal{H}t})$ down to basic single- and two-qubit gates. In practice, this is usually realized by Trotterization~\cite{trotter1959product}. Suppose $\mathcal{H} = \sum_j w_jP_j$ where $w_j \in \mathbb{R}$ and $\{P_j\}$ are Pauli strings. $exp({i\mathcal{H}t})$ can be approximated by the following trotterization product formula: 
\begin{equation*}
    e^{i\mathcal{H}t}=e^{it\sum_jw_jP_j}=\left(\prod_je^{iw_jP_j\Delta t}\right)^{t/\Delta t}+O(t\Delta t)
\end{equation*}
Each term $exp(iw_jP_j\Delta t)=exp(i\lambda_jP_j)$ ($\lambda_j = w_j\Delta t$) is converted to basic quantum gates.

\begin{figure}[h]
    \centering
    \includegraphics[width=0.7\linewidth]{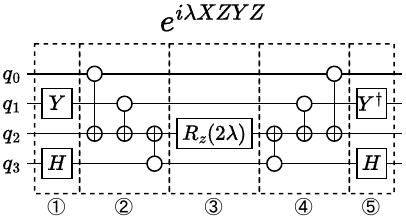}
    \caption{From Pauli string evolution operator $e^{i\lambda P}$ to corresponding circuit}
    \label{fig:circuit}
\end{figure}

Figure~\ref{fig:circuit} shows an example of how the Pauli string evolution operator $exp({i\lambda XZYZ})$ converts to its quantum circuit. It includes the following steps:
\begin{enumerate}
    \item[\textcircled{1}.] A layer of single-qubit gates is applied to each qubit, corresponding to its Pauli operator. A $H$ gate is applied if the corresponding operator is $X$ ($q_3$ in the example), and $Y$ is applied if the operator is $Y$ ($q_1$ in the example).
    \item[\textcircled{2}.] A target qubit ($q_2$ in the example) is selected. CNOT is applied to connect each qubit other than the target qubit whose corresponding Pauli operator is non-identity with the target qubit.
    \item[\textcircled{3}.] A $R_z(2\lambda)$ rotation is applied to the target qubit ($q_2$ in the example).
    \item[\textcircled{4}.] Apply the CNOT gates in \textcircled{2} reversely.
    \item[\textcircled{5}.] Apply the inverse single-qubit gates in \textcircled{1}. In this example, $Y^\dagger$ is applied to $q_1$ and $H$ to $q_3$.
\end{enumerate}

In general, when a qubit's corresponding operator is $I$ in the Pauli string $P$, then the circuit to implement the simulation of this Pauli string $e^{i\lambda P}$ will not result in any gates applied on that qubit. Only those qubits whose operators are Pauli operators will have gates involved. Roughly, the number of gates in the circuit implementation of $exp(i\lambda_jP_j)$ is proportional to the number of non-identity Pauli operators in the Pauli string $P_j$.

\subsubsection{Pauli Weight}
\label{sec:pauli-weight}

The Pauli weight of a Pauli string is defined by the number of non-identity Pauli operators in this string. For example, string $IIXX$ has a Pauli weight of 2. As discussed above, the Pauli weight is roughly proportional to the number of gates in the circuit implementation of $e^{i\lambda P}$~\cite{li2022paulihedral,sim-opt-degriend2020architectureaware}.
Therefore, a Hamiltonian $\mathcal{H}$ whose Pauli strings have the minimal sum of Pauli weight will have the minimal gate count when implementing $exp({i\mathcal{H}t})$ before any follow-up compilation/optimization.
Although today's quantum compiler involves many complex transformation and optimization passes, providing a good input circuit with minimal gate count for the downstream compilation/optimizations can, in general, benefit the final compiled circuit.


\subsubsection{Arithmetic}
\label{sec:pauli-string-arith}

Multiplying Pauli strings follows the rule of multiplication over tensor product: $P^1P^2=(\sigma^1_N \sigma^2_N)\otimes(\sigma^1_{N-1}\sigma^2_{N-1})\otimes\dots\otimes(\sigma^1_1\sigma^2_1)$, which is the tensor product of multiplying their corresponding Pauli operators.

With multiplication, we could define the anticommutator of two Pauli strings as $\{P_1, P_2\}=P_1P_2+P_2P_1$. Two Pauli strings \textit{anticommute} if their anticommutator is 0.

\subsection{Fermionic System}
\label{sec:fermionic-system}

The Fermionic quantum system refers to physical systems composed of Fermions. Typical Fermions include protons, electrons, and neutrinos. In the digital quantum simulation of Fermionic systems, the quantum state of Fermions is usually characterized by the occupation of Fermionic modes.
Since Fermions satisfy the Pauli exclusion principle, each Fermionic mode can either be unoccupied (denoted by the $\ket{0}_\mathcal{F}$) or occupied by at most one Fermion (denoted by the $\ket{1}_\mathcal{F}$).
Thus, each Fermionic mode has a 2-D state space $span\{\ket{0}_\mathcal{F}, \ket{1}_\mathcal{F} \}$. This is similar to a qubit but fundamentally different in statistical properties.
For example, exchanging the indices of two Fermionic modes will negate the state vector, but exchanging the indices of two qubits will not. Thus, we cannot directly map a Fermionic mode into a qubit when simulating a Fermionic system on a quantum computer, and a special Fermion-to-qubit encoding is required.

\subsubsection{Description of Fermionic Systems}
To understand the Fermion-to-qubit encoding, we first introduce the description of Fermionic systems. The basic operators to describe a Fermionic system with $N$ Fermionic modes are the $N$ creation and $N$ annihilation operators: $\{a_i^\dagger\}$, $\{a_i\}$ where $i=1\dots N$. These operators act on states that are described by vectors in the Fock space $\mathcal{F}(\mathbb{C}^N)$ (the state space of $N$ Fermionic modes), a $2^N$-D Hilbert space spanned by a set of orthonormal basis (Fock basis):
\begin{equation*}
    \ket{x_1,x_2,\dots x_N}_\mathcal{F}, \text{where}\ x_i=1\ \text{or}\ 0
\end{equation*}

Each $x_i$ is the occupation number of the corresponding mode $i$, also given by the occupation number operator $a_i^\dagger a_i$. The $2N$ creation and annihilation operators act on the basis vectors as:
$$   
    a_i^\dagger\ket{\dots0_i\dots}_\mathcal{F}=\ket{\dots1_i\dots}_\mathcal{F},  a_i\ket{\dots1_i\dots}_\mathcal{F}=\ket{\dots0_i\dots}_\mathcal{F}
$$
There exists a vacuum state $\ket{vac}_\mathcal{F}=\ket{0,\dots,0}_\mathcal{F}$ such that any annihilation operator applies on it results in 0:
$$
    \forall j,a_j\ket{vac}_\mathcal{F}=0
$$

The creation and annihilation operators in a Fermionic system must satisfy the Fermionic canonical anticommutativity:
\begin{equation}
    \begin{aligned}
        &\{a_i,a_j\}=\{a_i^\dagger,a_j^\dagger\}=0 \\
        &\{a_i^\dagger,a_j\}=\mathbb{I}\delta_{ij}
    \end{aligned}
\end{equation}

Here, $\delta_{ij}=0$ when $i\neq j$ and $\delta_{ij}=\delta_{ii}=1$ when $i=j$. $\mathbb{I}$ is the identity operator.

The Hamiltonian of a Fermionic System is usually a Hermitian operator expressed by the addition of production of the Fermionic creation and annihilation operators. For example, a 2-Fermionic-mode Hamiltonian can be: $$\mathcal{H}_\mathcal{F} = h_1a_1^{\dagger}a_1 + h_2a_2^{\dagger}a_2 $$
where $h_1,h_2 \in \mathbb{R}$ are parameters.

\subsubsection{Encoding Fermionic System in Qubit System}
To encode a Fermionic system on a quantum computer composed of qubits, we need to find a set of operators in the qubits state space that also satisfy the Fermionic canonical anticommutativity mentioned above.
This is usually achieved by finding Pauli strings for the so-called \textit{Majorana operators}, which can later be converted to the Fermionic creation and annihilation operators.

\textbf{Majorana operators:} The $N$ creation and $N$ annihilation operators could be paired into $2N$ Majorana operators to simplify the problem:
\begin{gather*}
    M_{2j}=a_j^\dagger+a_j\quad M_{2j-1}=i(a_j^\dagger-a_j) \\
    \Rightarrow\{M_i,M_j\}=2\mathbb{I}\delta_{ij}
\end{gather*}

Majorana operators are usually set to be Pauli strings. A simple example is the 2 Fermonic-mode system ($N=2$). With Jordan-Wigner transformation \cite{jordan_uber_1928}, a widely used Fermion-to-qubit encoding, the 4 Majorana operators are the following 4 Pauli strings:
\begin{equation}
    \begin{aligned}
        & M_1\mapsto IY\ && M_2\mapsto IX \\
        & M_3\mapsto YZ\ && M_4\mapsto XZ
    \end{aligned}
\end{equation}

Correspondingly:
\begin{equation*}
    \begin{aligned}
        & a_1^\dagger\mapsto0.5\cdot IX - 0.5i\cdot IY\ && a_1\mapsto0.5\cdot IX+0.5i\cdot IY \\
        & a_2^\dagger\mapsto0.5\cdot XZ-0.5i\cdot YZ\ && a_2\mapsto0.5\cdot XZ + 0.5i\cdot YZ
    \end{aligned}
\end{equation*}

The anticommutativity could be tested easily, given $\{a^\dagger_1,a_1\}$ as an example:
$$
    \begin{aligned}
        &\{a^\dagger_1,a_1\}=\{0.5\cdot IX - 0.5i\cdot IY,0.5\cdot IX+0.5i\cdot IY\} \\
        &=0.25\cdot(\{IX,IX\}-i\{IY,IX\}+i\{IX,IY\}+\{IY,IY\}) \\
        &=0.5\cdot II-0+0+0.5\cdot II = \mathbb{I}
    \end{aligned}
$$

Using this encoding, the Fermionic Hamiltonian example in the last section can be converted to a qubits system Hamiltonian in the following:
$$\mathcal{H}_\mathcal{F} = h_1a_1^{\dagger}a_1 + h_2a_2^{\dagger}a_2 $$
$$\mapsto \mathcal{H}_{qubit} = \frac{h_1+h_2}{2}\cdot II-\frac{h_1}{2}\cdot IZ-\frac{h_2}{2}\cdot ZI$$

\section{Fermion-to-Qubit Encoding via SAT}\label{sec:main-SAT-method}

The Fermion-to-qubit encoding is not unique. 
Different encodings will result in qubit Hamiltonians with different Pauli weights and circuit implementation overhead. 
In this section, we introduce \myFrameworkNameSpace to find the optimal Fermion-to-qubit encoding with minimal Pauli weight. We summarize the constraints and optimization objectives of finding a Fermion-to-qubit encoding and then introduce how they can be efficiently formulated into an SAT problem.

\subsection{Encoding Constraints and Objectives}
As introduced in Section~\ref{sec:fermionic-system},  the Fermionic creation and annihilation operators on $N$ Fermionic modes can be turned into $2N$ Majorana operators via a simple linear transformation, and the Majorana operators' constraints are much more straightforward. As a result, finding a Fermion-to-qubit encoding usually involves finding the Majorana operators and then pairing them to generate the Fermionic operators.

\textbf{Constraints:}\label{sec:constraints}
In summary, the $2N$ Majorana operators for an $N$-Fermion to $N$-qubit encoding are  $2N$ Pauli strings $\{S\}$ satisfying the following four constraints~\cite{RevModPhys.92.015003,raman_handbook_2000}: 
\begin{itemize}
    \item \textbf{Anticommutativity}: Any two of the $2N$ Majorana operators must anticommute.
    \begin{equation}
        \forall P_i,P_j\in\{S\},\{P_i,P_j\}=P_iP_j + P_jP_i=2\delta_{ij}
    \end{equation}
    \item \textbf{Linear independence}: All the $2N$ Majorana operators must be linear independent.
    \begin{equation}
        \sum_{i=1}^{2N}\alpha_iP_i = 0 \implies\alpha_i = 0, 1\leq i \leq 2N
    \end{equation}
    \item \textbf{Algebraic independence}: All the $2N$ Majorana operators must be algebraically independent. For any two unequal subsets of $\{S\}$, the multiplication of all the Pauli strings in one subset cannot be proportional to the multiplication of all Pauli strings in the other subset.
    \begin{equation}
        \forall S_a,S_b\subseteq\{S\},S_a\neq S_b\implies\prod_{P_a \in S_a} P_a\not\propto\prod_{P_b \in S_b} P_b
    \end{equation}
    \item \textbf{Vacuum state preserving}: The vacuum state $\ket{vac}_\mathcal{F}$ of Fock basis is represented by qubit state $\ket{0}^{\otimes N}$. This restricts the Majorana operators:
    \begin{equation}
        \forall 1\leq j\leq N,\frac{M_{2j}+iM_{2j+1}}{2}\ket{0}^{\otimes N}=0
    \end{equation}
    This constraint is optional and will not affect the correctness/optimality of a Fermion-to-qubit encoding.
\end{itemize}
Since the Pauli strings naturally formulate an orthonormal basis (see Section~\ref{sec:pauli-string-orthogonal}), satisfying the anticommutativity constraint, which requires all the Pauli strings to be different, already implies that the linear independence constraint is satisfied.
Consequently, the linear independence constraint is safely disregarded in the rest of this paper.

\textbf{Objective:}
Overall, the optimization objective of a Fermion-to-qubit encoding is to minimize the overhead of simulating this Fermionic system on the quantum computer, that is, the Pauli weight discussed in Section~\ref{sec:pauli-weight}.
In this paper, we adopt two types of objectives:
\begin{itemize}
   \item \textbf{Hamiltonian-independent:} We only consider the overhead of implementing the $2N$ Majorana operators. 
    The sum of the Pauli weights of all the $2N$ Majorana operators is minimized.
    These $2N$ Majorana operators are then used to generate the actual Hamiltonian.
    This solution may not be optimal for a specific Hamiltonian but can generally demonstrate good performance. 
    \item \textbf{Hamiltonian-dependent:} We encode the overhead of implementing the actual Hamiltonian of the target Fermionic system in the SAT optimization. This objective will give us the \textit{optimal} Fermion-to-qubit encoding that can achieve minimal implementation overhead for a specific Hamiltonian.
\end{itemize}

\subsection{Encode Majorana Operators}
Majorana operators are Pauli strings, while the SAT problem is formulated with Boolean variables.
Our first step is to encode the Pauli operators with Boolean variables and then map the operation of the Pauli operators/strings to Boolean expressions.

\textbf{Pauli Operator Encoding:} A Pauli operator $\sigma$  has four different possible values $\sigma\in\{X, Y, Z, I\}$ and can be encoded by a pair of two Boolean variables (two bits).
We denote the encoding by $E$:
$$
    E:\sigma\mapsto\{(\false,\false),(\false,\true),(\true,\false),(\true,\true)\}
$$
A possible encoding strategy $E$ is shown: 
\begin{align}\label{eqn:pauli-op-encoding}
    \begin{aligned}
        &E(I)=(\false, \false)&E(X)=(\false, \true) \\
        &E(Y)=(\true, \false)&E(Z)=(\true, \true)
    \end{aligned}
\end{align}

\textbf{Pauli Operator Multiplication Encoding:} Given the specific encoding strategy, multiplication between Pauli operators can then be expressed by Boolean expressions. 
The multiplication of two Pauli operators $\sigma_3=\sigma_1\sigma_2$ is summarized in Table~\ref{tab:pauli-op-mult}.
\begin{table}[t]
    \centering
    \caption{Truth table of Pauli operator multiplication $\sigma_3=\sigma_1\sigma_2$ using the Pauli operator encoding in Equation (\ref{eqn:pauli-op-encoding})}
    \begin{tabular}{|c|c|c|c|c|}\hline
         \diagbox{$\sigma_1$}{$\sigma_3$}{$\sigma_2$} & $I (\false,\false)$ & $X (\false,\true)$   & $Y (\true,\false)$   & $Z (\true,\true)$  \\\hline
        $I (\false,\false)$ & $I (\false,\false)$ & $X (\false,\true)$   & $Y (\true,\false)$   & $Z (\true,\true)$  \\\hline
        $X (\false,\true)$ & $X (\false,\true)$ & $I (\false,\false)$  & $iZ (\true,\true)$  & $-iY (\true,\false)$ \\\hline
        $Y (\true,\false)$ & $Y (\true,\false)$ & $-iZ (\true,\true)$ & $I (\false,\false)$   & $iX (\false,\true)$ \\\hline
        $Z (\true,\true)$ & $Z (\true,\true)$ & $iY (\true,\false)$  & $-iX (\false,\true)$ & $I (\false,\false)$  \\\hline
    \end{tabular}
    \label{tab:pauli-op-mult}
\end{table}
The additional coefficients produced by the multiplication can be ignored because they do not affect the algebraic independence and the anticommutativity checking. As a result, Table~\ref{tab:pauli-op-mult} can be considered the truth table when we derive the Boolean expression for the multiplication of the Pauli operator. 
The Boolean expression for $\sigma_3=\sigma_1\sigma_2$ is:
\begin{equation}\label{eqn:pauli-op-mult}
    \begin{aligned}
        E(\sigma_3).\mathit{1}=E(\sigma_1).\mathit{1}\oplus E(\sigma_2).\mathit{1}\\
        E(\sigma_3).\mathit{2}=E(\sigma_1).\mathit{2}\oplus E(\sigma_2).\mathit{2}
    \end{aligned}
\end{equation}
where $E(\sigma).\mathit{1}$ and $E(\sigma).\mathit{2}$ denote the first and the second bit encoding the Pauli operator $\sigma$. 

Note that this result indicates that multiplication is symmetric regarding the permutation of $X$, $Y$, $Z$, and $I$. Thus, any shifting of the encoding strategy would always produce a similar outcome without possessing unique algebraic properties, meaning the encoding scheme we selected in Equation~\eqref{eqn:pauli-op-encoding} does not lose generality.

\textbf{Pauli String Encoding:} We then extend the Boolean variable encoding to a Pauli string from an individual Pauli operator. A Pauli string of length $N$, $P=[\sigma_1,\dots,\sigma_N]$, has two equivalent forms of representation used in this paper:
\begin{itemize}
    \item The operator form:
    $$
        E_{op}(P)_i=E(\sigma_i)
    $$
    \item The bit sequence form:
    $$
        E_{bit}(P)_i=\left\{\begin{aligned}
            &E(\sigma_{(i+1)/2}).\mathit{1}&i\ \mathrm{is}\ \mathrm{odd}\\
            &E(\sigma_{i/2}).\mathit{2}&\mathrm{otherwise}
        \end{aligned}\right.
    $$
\end{itemize}

\textbf{Pauli String Multiplication Encoding:} The Boolean expression for the multiplication of two Pauli strings can be the combination of the Pauli operator multiplication at each location, following the definition in Section~\ref{sec:pauli-string-arith}.
Suppose $P^1=[\sigma^1_1,\dots,\sigma^1_N]$ and $P^2=[\sigma^2_1,\dots,\sigma^2_N]$. Then $$P^1P^2 = [\sigma^1_1\sigma^2_1,\dots,\sigma^1_N\sigma^2_N],\ E_{op}(P^1P^2)_i = E(\sigma^1_i\sigma^2_i)$$

\subsection{Anticommutativity Constraints}
To encode the anticommutativity constraint, we first encode the anticommutativity of Pauli operators and then extend to Pauli strings.

\textbf{Anticommutativity of Pauli Operators:} 
Suppose we use $0$ and $1$ to denote that two Pauli operators $\sigma_1$ and $\sigma_2$ are anticommute or not, respectively, and $\facomm(E(\sigma_1),E(\sigma_2))$ to denote the Boolean expression to determine the anticommutativity of $\sigma_1$ and $\sigma_2$. 
Table~\ref{tab:pauli-op-acomm} shows the truth table of $\facomm(E(\sigma_1), E(\sigma_2))$. 
Notice that $I$ does not anticommute with any operator. Then the Boolean expression of $\facomm(E(\sigma_1),E(\sigma_2))$ is shown in the following: 
\begin{table}[t]
    \centering
    \caption{Anticommutativity of Pauli Operators}
    \begin{tabular}{|c|c|c|c|c|}\hline
        \diagbox{$\sigma_2$}{$\sigma_1$} & $I (\false,\false) $ & $X (\false,\true)$   & $Y (\true,\false)$   & $Z (\true,\true)$  \\\hline
        $I (\false,\false)$ & $\false$ & $\false$ & $\false$ & $\false$ \\\hline
        $X (\false,\true)$ & $\false$ & $\false$ & $\true$ & $\true$ \\\hline
        $Y (\true,\false)$ & $\false$ & $\true$ & $\false$ & $\true$ \\\hline
        $Z (\true,\true)$ & $\false$ & $\true$ & $\true$ & $\false$  \\\hline
    \end{tabular}
    \label{tab:pauli-op-acomm}
\end{table}
\begin{equation}\label{eqn:pauli-op-acomm}
    \begin{aligned}
        \facomm(E(\sigma_1),&E(\sigma_2))=\\
        &(E(\sigma_1).\mathit{1}\land\lnot E(\sigma_2).\mathit{1}\land E(\sigma_2).\mathit{2})\lor\\
        &(E(\sigma_1).\mathit{2}\land E(\sigma_2).\mathit{1}\land\lnot E(\sigma_2).\mathit{2})\lor\\
        &(E(\sigma_2).\mathit{1}\land\lnot E(\sigma_1).\mathit{1}\land E(\sigma_1).\mathit{2})\lor\\
        &(E(\sigma_2).\mathit{2}\land E(\sigma_1).\mathit{1}\land\lnot E(\sigma_1).\mathit{2})
    \end{aligned}
\end{equation}

\textbf{Anticommutativity of Pauli Strings:} 
Note that for any two Pauli operators, they either commute or anticommute:
$$\sigma_i\sigma_2 = (+1)\sigma_2\sigma_1\ \text{or}\ \ \sigma_1\sigma_2 = (-1)\sigma_2\sigma_1 $$
Then, in the anticommute check for two Pauli strings,
each pair of anticommute Pauli operators will introduce one factor of $(-1)$ in the anticommutivity check of two Pauli strings:
\begin{align*}
    P^1P^2 + P^2P^1 = & [\sigma^1_1\sigma^2_1,\dots,\sigma^1_N\sigma^2_N] +  [\sigma^2_1\sigma^1_1,\dots,\sigma^2_N\sigma^1_N] \\
    & [\sigma^1_1\sigma^2_1,\dots,\sigma^1_N\sigma^2_N] + (-1)^a[\sigma^1_1\sigma^2_1,\dots,\sigma^1_N\sigma^2_N] 
\end{align*}
where $a$ is the number of anticommute Pauli operator pairs $\sigma^1_k$ and $\sigma^2_k$, $1\leq k \leq N$.
The anticommutativity between two distinct Majorana operators $P_i$ and $P_j$ (whether $P_iP_j + P_iP_j = 0$) is equivalent to if they have an odd number $a$ of qubits whose Pauli operators anticommute. 
For example, the Pauli operator $X$ and $Y$ anticommute. Then, two Pauli strings $XX$ and $YY$ will not anticommute because they share \textit{two} pairs of anticommute operators $X$ and $Y$. While the Pauli strings $XXX$ and $YYY$ will anticommute because they share \textit{three} pairs of anticommute Pauli operators.
This principle is equivalent to all the $\facomm(E_{op}(P_i)_k,E_{op}(P_j)_k)$ of index $1\le k\le N$ xor to $\true$:
$$
    \bigoplus_{k=1}^N\facomm(E_{op}(P_i)_k,E_{op}(P_j)_k)=\true
$$

Since the problem requires all possible pairs of $2N$ Majorana operators to be anticommute, the model $\mathcal{M}$ of a valid Fermion-to-qubit encoding should satisfy the conjunction of all anticommutativity of Pauli string pairs:
$$
    \mathcal{M}\models\bigwedge_{1\leq i<j\leq 2N}\bigoplus_{k=1}^N\facomm(E_{op}(P_i)_k,E_{op}(P_j)_k)
$$

\subsection{Algebraic Independence Constraints}\label{sec:algebraic-independence-constraints}
We first analyze the algebraic independence condition.
Suppose we have two sets of Pauli strings $S_a$ and $S_b$, $S_a \neq S_b$. 
If the algebraic independence condition is violated, we will have $\prod_{P_a \in S_a} P_a\propto\prod_{P_b \in S_b} P_b$.
We can ignore the coefficients in the Pauli string multiplication, and then the condition becomes equivalence checking of the multiplication results:
$$\prod_{P_a \in S_a} P_a = \prod_{P_b \in S_b} P_b \iff \prod_{P_a \in S_a} P_a\prod_{P_b \in S_b} P_b = I_1I_2\dots I_N$$
Two Pauli strings are equal if and only if their multiplication is the all-identity string.
Note that if $S_a$ and $S_b$ share some Pauli strings, removing those shared Pauli strings from the two sets will not change this result.  
Consequently, we only need to consider disjoint sets $S_a$ and  $S_b$ where $S_a \cap S_b =  \varnothing$.
In this case, the condition is turned into:
$$\prod_{P \in S_a \cup S_b} P  =  I_1I_2\dots I_N$$
Here $S_a \cup S_b$ can be an arbitrary subset of the $2N$ Majorana operators set, denoted as $S_*$.

For certain $S_*$, the deduced algebraic \textit{dependence} could be tested via the equality in bit sequence form. The left side follows the encoding of operator multiplication, which is the xor of all corresponding bits:
$$
    E_{bit}\left(\prod_{P\in S_*} P\right)=\bigoplus_{P\in S_*} E_{bit}(P)
$$

The bit sequence of $I$ implies that all bits are $\false$, which is enforced by:
\begin{equation}\label{eqn:bit-sequence-identity}
    \bigwedge_{j=1}^{N}\lnot\left(\bigoplus_{P\in S_*} E_{bit}(P)_j\right)=\true
\end{equation}

The encoding model $\mathcal{M}$ should not allow any algebraic \textit{dependence} for $S_*\subseteq S$. Thus, for any $S_*$ in the power set of $S$ (denoted as $\mathcal{P}(S)$), logic not of Equation~\eqref{eqn:bit-sequence-identity} must be satisfied:
$$
    \begin{aligned}
        \mathcal{M}\models&\bigwedge_{S_*\in\mathcal{P}(S)}\lnot\bigwedge_{j=1}^{N}\lnot\left(\bigoplus_{P\in S_*} E_{bit}(P)_j\right) \\
        =&\bigwedge_{S_*\in\mathcal{P}(S)}\bigvee_{j=1}^{N}\left(\bigoplus_{P\in S_*} E_{bit}(P)_j\right)
    \end{aligned}
$$
Since $S$ has $2N$ elements, its power set $\mathcal{P}(S)$ has $2^{2N}$ elements.
The constraint of algebraic independence thus incurs a large overhead by generating $2^{2N}$ clauses.
We will later show how to reduce such complexity in Section~\ref{sec:ignoring-algebraic-independence}.

\subsection{Vacuum State Preservation}

Vacuum state preservation requires the encoding maps the vacuum state $\ket{vac}_\mathcal{F}$ to $\ket{0}^{\otimes N}$. It sets requirements for the mapped annihilation operators:
\begin{equation}\label{eqn:vacuum-state-preservation}
    a_j\ket{vac}_\mathcal{F}=0\implies\frac{M_{2j}+iM_{2j+1}}{2}\ket{0}^{\otimes N}=0
\end{equation}

A simple case where Equation~\eqref{eqn:vacuum-state-preservation} holds is when there exists at least one index $k$ such that $(M_{2j})_k+i(M_{2j+1})_k=0$, which indicates that at such index $k$, the Pauli operators $(M_{2j})_k$ and $(M_{2j+1})_k$, is a pair of $X$ and $Y$. This property could be imposed by assuming the final solution produced by the SAT solver is correctly paired; that is, $M_k=P_k$. The creation and annihilation operators are:
\begin{equation}\label{eqn:majorana-to-operator}
    a_j^\dagger=\frac{P_{2j}-iP_{2j-1}}{2},a_j=\frac{P_{2j}+iP_{2j-1}}{2}
\end{equation}

The above condition could thus be encoded via Boolean constraints. The existence of $XY$ pair between Pauli operator $\sigma_1$ and $\sigma_2$ could be tested by the following in our encoding:
$$
    \begin{aligned}
        \mathrm{pair}&(E(\sigma_1),E(\sigma_2))=\\&\lnot E(\sigma_1).\mathit{1}\land E(\sigma_1).\mathit{2}\land E(\sigma_2).\mathit{1}\land\lnot E(\sigma_2).\mathit{2}
    \end{aligned}
$$

A clause is generated for each Majorana operator pair $P_{2j}$ and $P_{2j+1}$, which should have at least one $XY$ pair across all possible indexes. The final encoding model $\mathcal{M}$ is expected to satisfy each Majorana operator pair $P_{2j},P_{2j+1}$ for $j=1$ to $N$:
$$
    \mathcal{M}\models\bigwedge_{j=1}^N\bigvee_{k=1}^N\mathrm{pair}(E_{op}(P_{2j})_k,E_{op}(P_{2j+1})_k)
$$


\subsection{Hamiltonian-Independent Weight Constraint}\label{sec:maximum-weight-constraint}
A widely used Hamiltonian-independent optimization objective is to minimize the sum of the Pauli weights of all the $2N$ Majorana operators.
Recall the Pauli weight of a Pauli string is the number of non-identity Pauli operators in the string (Section~\ref{sec:pauli-weight}).
A single Pauli operator will contribute to the total Pauli weight if and only if it is not an identity. Using our Pauli operator encoding in Equation~\eqref{eqn:pauli-op-encoding}, the weight of a Pauli operator is: 
$$
        \weight(E(\sigma))  = 
        E(\sigma).\mathit{1}\lor E(\sigma).\mathit{2} 
$$

The total weight then accumulates the weight of each single operator, and this is our optimization target:
$$
    \min_{P_1, \dots, P_{2N}}\sum_{k=1}^{2N}\sum_{i=1}^N \weight(E_{op}(P_k)_i)
$$
subject to all the constraints mentioned earlier in this section.

Note that a SAT solver cannot automatically optimize for such a target. In practice, we set a maximum Pauli weight of $w$ 
and encode this as a constraint in the model:
$$
    \sum_{k=1}^{2N}\sum_{i=1}^N \weight(E_{op}(P_k)_i) < w
$$
An SAT solver can determine whether a solution exists for a given $w$. We start from a larger $w$ and gradually reduce $w$ until the SAT solver cannot find a satisfactory solution in a given time. This will give the minimal $w$ where a valid Fermion-to-qubit encoding could exist. The overall procedure is summarized in Algorithm~\ref{alg:minimum-pauli-weight}.
\begin{algorithm}[t]
    \caption{Solve $\mathcal{M}$ with Optimal Pauli Weight}
    \label{alg:minimum-pauli-weight}
    \begin{algorithmic}
        \REQUIRE $C:\mathrm{Clauses}$
        \REQUIRE $\mathcal{M}:\mathrm{Model}\models C$
        \ENSURE $\mathcal{M}:\mathrm{Model}\models C\land\min w$
        \WHILE{$\mathcal{M}'\gets\mathrm{solve}(C) == \mathrm{SAT}$}
            \STATE $\mathcal{M}\gets\mathcal{M}'$
            \STATE $S=\{P_k\}\gets\mathrm{decode}(\mathcal{M})$
            \STATE $w\gets\sum_{k=1\dots2N}\sum_{i=1\dots N}\weight(E_{op}(P_k)_i)$
            \STATE $C\gets\sum_{k=1\dots2N}\sum_{i=1\dots N}\weight(E_{op}(P'_k)_i)<w\land C$
        \ENDWHILE
        \RETURN $\mathcal{M}$
    \end{algorithmic}
\end{algorithm}

To obtain the initial feasible solution, we have to start with a total weight $w_0$ such that it could produce a solution but also close enough to the minimum weight to reduce the solving time. In practice, we start from the Pauli weight of the baseline Bravyi-Kitaev transformation~\cite{bravyi_fermionic_2002}.

\subsection{Hamiltonian-Dependent Weight Constraint}
\label{sec:hamiltonian-pauli-weight}
The Hamiltonian-independent weight constraint can find the set of Majorana operators with minimal Pauli weight.
Generally, we can implement the Hamiltonian simulation using these Majorana operators with relatively small overhead.
However, this is still not the optimal Pauli weight for a specific Hamiltonian because different Majorana operators are multiplied and added in different ways with real Hamiltonians.
For example, a molecular electron Hamiltonian has the following form:
\begin{equation}\label{eqn:system-hamiltonian}
    \mathit{H}=\sum_{i,j} h_{ij}a^\dagger_ia_j+\frac{1}{2}\sum_{i,j,k,l} h_{ijkl}a^\dagger_ia^\dagger_ja_ka_l
\end{equation}
The multiplication $a_i^\dagger a_j$ and $a^\dagger_ia^\dagger_ia_ka_l$ makes it possible to cancel out Pauli operators at specific indices.
This is not considered in the Hamiltonian-independent weight constraint. 
To achieve the optimal Pauli weight for a specific Hamiltonian, the Fermion-to-qubit encoding should consider the structure of the targeted Hamiltonian and encode it in the SAT problem.

We first find the Boolean expression for the actual total weight of a Hamiltonian.
A Hamiltonian is in the form of the summation of the multiplication of the creation and annihilation operators.
The multiplication of several creation and annihilation operators is denoted as:
$$
    a_i^{(\dagger)}a_j^{(\dagger)}a_k^{(\dagger)}\dots, \text{where}\ i\neq j\neq k\neq\dots
$$
Each creation/annihilation operator would be decomposed into two Majorana operators $M_{2i}$ and $M_{2i-1}$ (Equation~\ref{eqn:majorana-to-operator}). Thus, the total Pauli weight of such multiplication would be calculated as:
\begin{equation}\label{eqn:hamiltonian-pauli-weight}
    \begin{aligned}
        &\weight(a_i^{(\dagger)}a_j^{(\dagger)}a_k^{(\dagger)}\dots)=\\
        &\sum_{x=2i,2i-1;y=2j,2j-1;z=2k,2k-1;\dots}\weight(M_xM_yM_z\dots)
    \end{aligned}
\end{equation}
This Hamiltonian-dependent weight constraint encodes the total Pauli weight of implementing a specific Hamiltonian. We can use this new weight constraint in the SAT formulation to find the optimal encoding for a targeted Hamiltonian. The other constraints and overall structure of Algorithm~\ref{alg:minimum-pauli-weight} remain the same.

\subsection{CNF Conversion and Solving}

Our entire encoding framework only uses Boolean variables with no natural number theory.
This allows us to transform the weight constraints into pure Boolean expressions. Along with other conditions, pure SAT solvers could solve the problem entirely, which would typically be dramatically faster than solving the SMT version of the problem.
The last step towards a pure SAT problem is to convert clauses into the Conjunctive Normal Form. Directly unfolding $xor$'s would cause an exponential explosion in clause size and be unbearable even on a small scale. Here, we adopt the Tseitin transformation~\cite{Tseitin1983} to solve such conversion in linear time of number of clauses by introducing additional Boolean variables polynomial in the size of the original formula.
For a Hamiltonian with $N$ Fermionic modes,
the number of variables is $\mathbf{O(4N^2)}$ and the number of clauses is $\mathbf{O(4^N)}$.

\section{Scaling up the SAT Method}


As discussed above, the number of clauses grows exponentially as the number of Fermionic modes increases, which becomes the immediate bottleneck of our framework.
To solve larger-scale Fermion-to-qubit encodings, we identify two optimization opportunities and propose corresponding techniques to reduce the number of clauses.

\subsection{Ignoring Algebraic Independence}
\label{sec:ignoring-algebraic-independence}
The first cause of the large number of clauses is the algebraic independence constraints (Section~\ref{sec:algebraic-independence-constraints}).
To deal with this overhead, we made the following key observations:

\textit{The constraint of algebraic independence could be ignored when $N$ is large.}

\begin{figure}[t]
    \centering
    \includegraphics[width=0.65\linewidth]{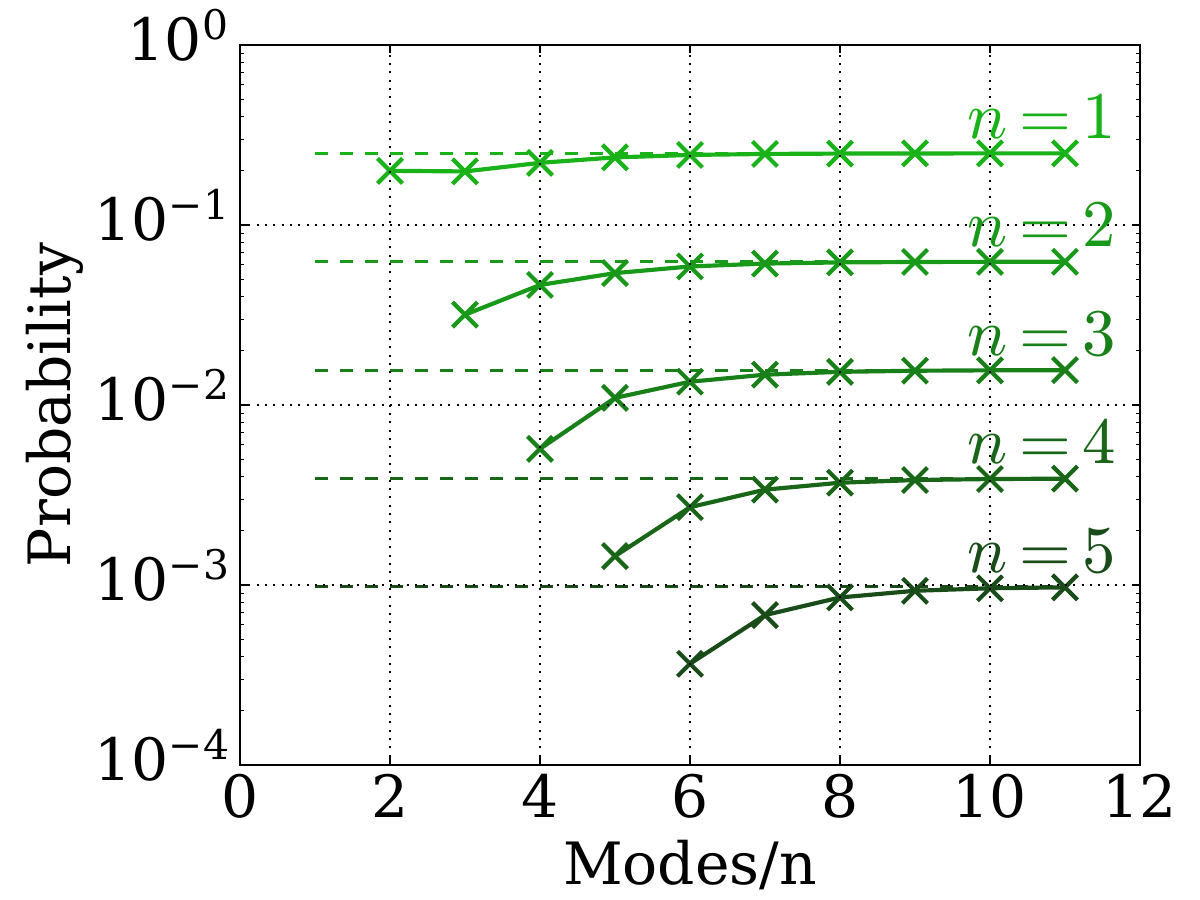}
    \caption{Probability of $n$ $A_k$'s holds simultaneously}
    \label{fig:probability-hold}
\end{figure}

We find that the probability for our SAT-based framework to generate an invalid encoding without considering the algebraic independence constraints is only $\frac{1}{4^N}$, where $N$ is the number of Fermionic modes.
To understand this, we define solutions found under reduced constraints as \textit{approximate solutions}. Consider the probability that a subset of Majorana operators forms an algebraic dependence. It means all corresponding Pauli operators at each index $k$ multiply to $I$ (coefficient ignored):
$$
    \forall 1\leq k\leq N, \prod (M_i)_k=I
$$
\begin{equation}\label{eqn:probability-single-k}
    A_k:\prod (M_i)_k=I
\end{equation}

For a certain $k$, Equation (\ref{eqn:probability-single-k}) holds when the quantity of $X$, $Y$, and $Z$ are all even or odd, which, the probability is approximately $1/4$ under the assumption that all Pauli operators are distributed uniformly and randomly.
Moreover, if the Pauli operator distributions are independent at each index, the probability that all the $A_k$'s hold simultaneously to break the algebraic independence constraints is $\frac{1}{4^N}$ for $N$ Fermionic modes.
This exponentially small failing probability indicates that when $N$ is significant, the $2N$ Majorana operators are most likely algebraic independent, and it would be safe to ignore such constraints to remove their exponentially large number of clauses. 

All the discussion above assumes that the Pauli operator distributions are independent at each index.
Figure~\ref{fig:probability-hold} shows numerical evidence for this assumption.
For small-scale cases with 1 to 11 Fermionic modes, we select the first 50 optimal Fermion-to-qubit encodings generated from our framework with the algebraic independence constraints applied.
We evaluate the probability of $n$ $A_k$'s holds simultaneously over the sampled optimal encodings.
The data in Figure~\ref{fig:probability-hold} perfectly matches our observation that the probability of  $n$ $A_k$'s holds simultaneously is $\frac{1}{4^n}$.
Therefore, we are confident that the failing probability after removing the algebraic independence constraints is $\frac{1}{4^N}$ for $N$ Fermionic modes.




\begin{algorithm}[t]
    \caption{Solve Hamiltonian Pauli Weight by \textbf{Annealing}}
    \label{alg:annealing}
    \begin{algorithmic}
        \REQUIRE $\mathcal{H}: N-\mathrm{mode\ Hamiltonian}$
        \REQUIRE $T_0: \mathrm{initial\ temperature}, T_1: \mathrm{final\ temperature}$
        \REQUIRE $\alpha: \mathrm{temperature\ step}$
        \REQUIRE $i: \mathrm{iterations}$
        \REQUIRE $m: 2N\ \mathrm{Majorana\ operators}$
        \ENSURE $2N\ \mathrm{Majorana\ operators}$
        \STATE $T\gets T_0$
        
        \STATE $w=pauli\_weight(m,\mathcal{H})$
        
        \WHILE{$T\geq T_1$}
            \FOR{$1\dots i$}
                \STATE $x,y=randint(1,N)$
                \STATE $swap(x, y, m)$
                \STATE $w'=pauli\_weight(m,\mathcal{H})$
                \IF{$random()\geq e^{-\frac{(w'-w)k}{T}}$}
                    \STATE $swap(x, y, m)$ /* undo the swap */
                \ENDIF
            \ENDFOR
            \STATE $T\gets T-\alpha$
        \ENDWHILE
        \RETURN $m$
    \end{algorithmic}
\end{algorithm}

\subsection{Simulated Annealing}
\label{sec:annealing}
We observe that the second cause of the large number of clauses is the Hamiltonian-dependent Pauli weight. Based on Equation~\eqref{eqn:system-hamiltonian}, second-quantization terms usually grow exponentially with Fermion modes. Our solution to this problem is to remove this part from the SAT formulation.
We can first find $2N$ Majorana operators with the Hamiltonian-independent Pauli weight, then find the pairing of the Majorana operators to the creation/annihilation operators.



We adopt the simulated annealing algorithm to solve this assignment problem, shown in Algorithm~\ref{alg:annealing}.
Given an $N$ mode Fermionic system, we could obtain the general Fermion-to-qubit encoding via the original method and try different sequences of Majorana operators to form the corresponding $N$ creation and $N$ annihilation operators. The annealing process takes the Hamiltonian Pauli weight as ``energy'', and two Majorana operators are swapped in each iteration. Several classic oracles used in the algorithm are:
\begin{itemize}
    \item $randint(i,j)$: return a random integer in $[i,j]$.
    \item $random()$: return a random real number in $[0,1]$.
    \item $pauli\_weight(m,\mathcal{H})$: calculates the Hamiltonian Pauli weight with given $2N$ strings, where the creation and annihilation operators are paired as Equation~\eqref{eqn:majorana-to-operator}.
    \item $swap(i,j,m)$: swap the $i^{th}$ and $j^{th}$ creation and annihilation operators, that is, swapping the $2i$ with $2j$ string and $2i+1$ with $2j+1$ string. This swap does not break the vacuum state preservation property since we are not breaking any existing pairing.
\end{itemize}

\subsection{Complexity Analysis}

The SAT problem of finding $N$-mode Fermion-to-qubit encoding requires $O(N^2)$ Boolean variables, $O(4^N)$ clauses for algebraic independence constraints, $O(N^2)$ clauses for anticommutativity constraints, and $O(N)$ clauses for vacuum state preservation. The number of clauses for Hamiltonian-dependent Pauli weight depends on the specific Hamiltonian. For the benchmark Hamiltonians used later in the Evaluation Section of this paper, the numbers of clauses for Hamiltonian-dependent Pauli weight are $O(N^4)$ for electronic structure problems~\cite{RevModPhys.92.015003} and SYK model~\cite{sykmodel}, and $O(N^2)$ for the Fermi-Hubbard model~\cite{hubbard}. Applying our two optimizations in Section 4 will reduce the number of clauses to $O(N^2)$.

To better understand the effect of eliminating the clauses for algebraic independence, Table~\ref{tab:clauses-variables-number} shows the intermediate data in the formulated SAT instances, including the number of variables, the number of clauses, and the average number of variables per clause w/ and w/o the algebraic independence constraints when using the Hamiltonian-independent weight constraint. (N/A denotes that generating the corresponding SAT instance takes over 1 hour.) It can be observed that eliminating algebraic independence can significantly simplify the generated SAT instance by reducing the number of variables and clauses.

\begin{table}[t]
    \centering
     \caption{Number of variables and clauses w/ and w/o algebraic independence in the generated SAT instances}
    \resizebox{\linewidth}{!}{\begin{tabular}{|c|c|c|c|c|c|c|}\hline
         \multirowcell{2}{Fermionic\\ Modes $N$} & \multicolumn{2}{c|}{\#Vars} & \multicolumn{2}{c|}{\#Clauses} & \multicolumn{2}{c|}{\makecell{Average\\ \#Vars/\#Clauses}} \\\cline{2-7}
         & w/ & w/o & w/ & w/o & w/ & w/o \\\hline
         2 & 70 & 46 & 459 & 331 & 3.65 & 3.72 \\\hline
         3 & 417 & 129 & 2436 & 1147 & 3.58 & 3.72 \\\hline
         4 & 2224 & 352 & 10926 & 3014 & 3.41 & 3.98 \\\hline
         5 & 10570 & 610 & 46925 & 5801 & 3.29 & 4.03 \\\hline
         6 & 49902 & 1158 & 210064 & 10601 & 3.23 & 4.02 \\\hline
         7 & 230503 & 1687 & 948732 & 16608 & 3.21 & 4.05 \\\hline
         8 & 1050544 & 2704 & 4283375 & 25693 & 3.21 & 4.04 \\\hline
         9 & N/A & 3600 & N/A & 36037 & N/A & 4.06 \\\hline
         10 & N/A & 5230 & N/A & 50798 & N/A & 4.05 \\\hline
         11 & N/A & 6589 & N/A & 66593 & N/A & 4.06 \\\hline
         12 & N/A & 8976 & N/A & 88440 & N/A & 4.05 \\\hline
         13 & N/A & 10894 & N/A & 111129 & N/A & 4.06 \\\hline
         14 & N/A & 14182 & N/A & 141504 & N/A & 4.05 \\\hline
         15 & N/A & 16755 & N/A & 172132 & N/A & 4.06 \\\hline
         16 & N/A & 21088 & N/A & 211938 & N/A & 4.06 \\\hline
         17 & N/A & 24412 & N/A & 252025 & N/A & 4.06 \\\hline
         18 & N/A & 29934 & N/A & 302793 & N/A & 4.06 \\\hline
    \end{tabular}}
    \label{tab:clauses-variables-number}
\end{table}

\section{Evaluation}

This section evaluates our SAT-based optimal Fermion-to-qubit encoding on various benchmarks with compilation output, noisy simulation, and real system testing.


\subsection{Experiment Setup}

\textbf{Benchmark:}
First, we have evaluated Majorana operator sets of various sizes as the Hamiltonian-independent benchmark.
Then, we prepare three representative Hamiltonians from different domains, all of which are widely used in quantum simulation studies:
a) molecule electron structure problem from quantum chemistry~\cite{RevModPhys.92.015003}, b) 1-D Fermi-Hubbard model with periodic boundary condition from condensed matter physics~\cite{hubbard}, and c) four-body SYK model from quantum field theory~\cite{sykmodel}.
Their second quantized~\cite{dirac1927quantum,jordan1928uber} Hamiltonians, as introduced in Section~\ref{sec:fermionic-system}, are listed in Figure~\ref{fig:hamiltonians}. We generate the multiplication and summation structure of creation and annihilation operators, transform them into Majorana operators, and fit in the optimal Hamiltonian Pauli weight solver.
\begin{figure}[t]
    \centering \small
\begin{align*}
    \mathit{H}_{electronic}=\sum_{i,j}^N h_{ij}a^\dagger_ia_j+\frac{1}{2}\sum_{i,j,k,l}^N h_{ijkl}a^\dagger_ia^\dagger_ja_ka_l \\
\mathit{H}_{Fermi-Hubbard}=-t\sum_{\sigma\in\{\uparrow,\downarrow\}}\sum_{i,j}^{N/2}(a^\dagger_{i\sigma}a_{j\sigma}+a^\dagger_{j\sigma}a_{i\sigma})\\
+U\sum_i^{M/2}a^\dagger_{i\uparrow}a_{i\uparrow}a^\dagger_{i\downarrow}a_{i\downarrow} \\
\mathit{H}_{SYK}=\frac{1}{4\times4!}\sum_{i,j,k,l=1}^N g_{ijkl}M_iM_jM_kM_l
\end{align*}
    \caption{The three types of benchmark Hamiltonians used in this paper. $a^\dagger_*$ and $a_*$ are the creation and annihilation operators. $M_*$ is the Majorana operator.}
    \label{fig:hamiltonians}
\end{figure}


\textbf{Implementation:}
We implement the proposed SAT-based Fermion-to-qubit encoding and execute our experiments with the following key components:
\begin{itemize}
    \item \textbf{SAT Solver}: We use two solvers: 
    a) the Z3~\cite{z3-solver} solver for formulating constraints and applying the Tseitin transformation, and b) the Kissat~\cite{kissat-solver} solver as the standalone high-performance SAT solver. 
    Cadical~\cite{cadical} is a backup solver when Kissat behaves abnormally. The SAT solver runs on CPUs only. 
    \item \textbf{Simulation Platform}: We use the Qiskit Aer simulator~\cite{qiskit} to execute the noisy simulations. The SAT solver and the noisy simulator run on a server with one AMD EPYC CPU (96 cores, frequency 2.4 GHz) and 768 GB memory.
    \item \textbf{Real System Study}: We perform an actual system study on the IonQ Aria-1 quantum computer, available through Amazon Braket. This device has 25 fully connected ion-trap qubits. It has $99.99\%$ single-qubit gate fidelity, $98.91\%$ two-qubit gate fidelity, and a $98.82\%$ readout fidelity.
\end{itemize} 

\textbf{Baseline:}
Our baseline is the Jordan-Wigner (JW)~\cite{jordan_uber_1928} and the Bravyi-Kitaev (BK) transformation~\cite{bk-compare-with-jw} implemented by Qiskit~\cite{qiskit}.

\textbf{Metrics:}
We generally use the \textbf{Pauli Weight} to indicate the performance of a Fermion-to-qubit encoding. Also, we evaluate the \textbf{gate count} as another indicator of the quality of our Fermion-to-qubit encoding compilation. We use the \textbf{energy} of simulated states for noisy simulation and real system studies and compare it with theoretical results.

\textbf{Configurations:}
We have three experimental configurations: 1) \textbf{\textsf{Full SAT}} is encoding all applicable constraints in SAT (Section~\ref{sec:hamiltonian-pauli-weight}).
2) \textbf{\textsf{SAT w/o Alg.}} is to simplify the SAT solving by removing the algebraic constraints (Section~\ref{sec:ignoring-algebraic-independence}).
3) \textbf{\textsf{SAT + Anl.}} is further assigning the Majorana operators with simulated annealing rather than encoding Hamiltonian-dependent weight into the SAT (Section~\ref{sec:annealing}).

\subsection{Hamiltonian-Independent Encoding}

\begin{figure}[t]
    \centering
    \includegraphics[width=0.6\linewidth]{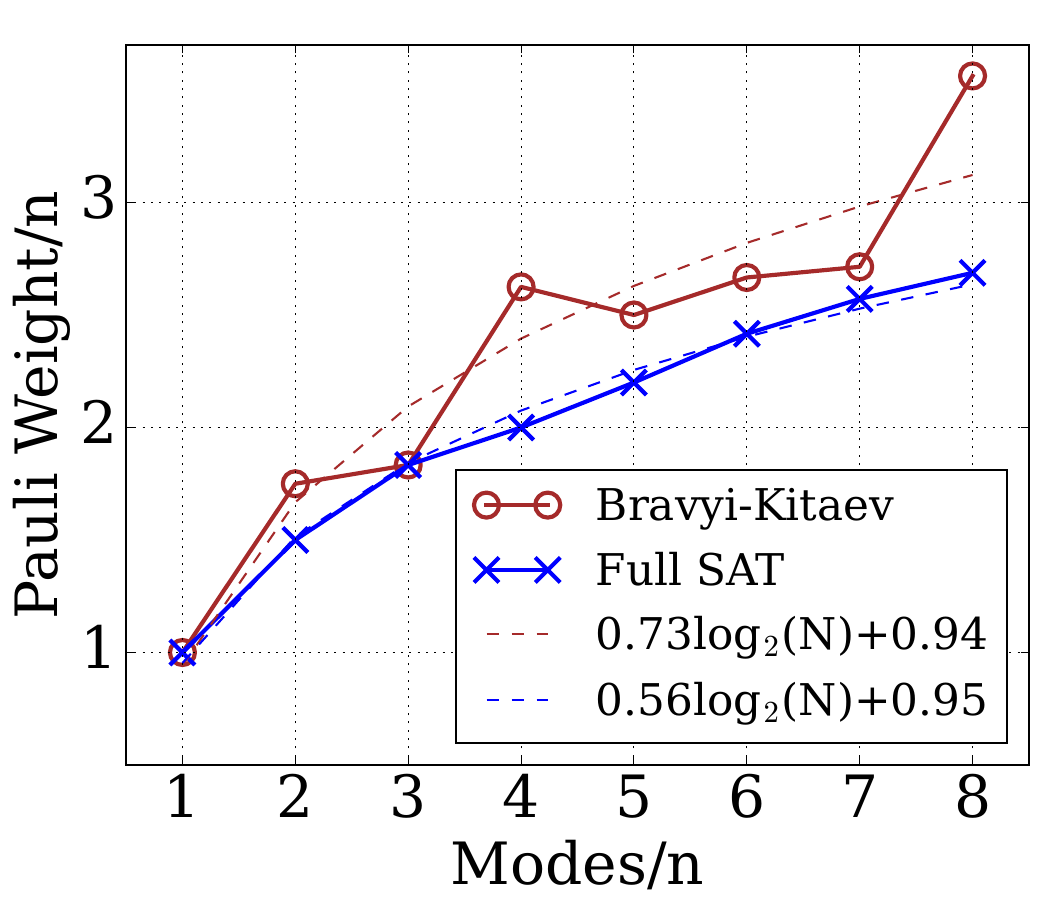}
    \caption{Average Pauli weight per-Majorana operator (small scale)}
    \label{fig:small-scale-average}
\end{figure}

In this section, we evaluate the performance of Hamiltonian-independent Fermion-to-qubit encoding. We first compare \textbf{\textsf{Full SAT}} with BK at a small scale (up to 8 Fermionic modes). Figure~\ref{fig:small-scale-average} shows the average Pauli weight per Majorana operator from 1 to 8 Fermionic modes. On average, our method shows a $11.16\%$ reduction in per-operator Pauli weight on a small scale. We also plot the regression of the data points.

\begin{table*}[t]
    \centering
    \caption{Hamiltonian-dependent total Pauli weight (small scale), BK vs. \textbf{\textsf{SAT+Anl.}}  vs. \textbf{\textsf{Full SAT}}}
    \begin{tabular}{|c|c|c|c|c|c|c|}\hline
        Case & Modes ($N$) & Bravyi-Kitaev & \textbf{\textsf{SAT+Anl.}} & Reduction & \textbf{\textsf{Full SAT}} & Reduction \\\hline
        \multirow{2}{*}{Electronic Structure} & \textit{4} & $934$ & $988$ & $-5.78\%$ & $790$ & $15.42\%$ \\\cline{2-7}
                                            & \textit{6} & $9004$ & $6878$ & $23.61\%$ & $6354$ & $29.43\%$ \\\Xhline{4\arrayrulewidth}
                                            
        \multirow{3}{*}{Fermi-Hubbard} & \textit{4} & $90$ & $104$ & $-15.56\%$ & $72$ & $20\%$ \\\cline{2-7}
                                        & \textit{6} & $284$ & $202$ & $28.87\%$ & $182$ & $35.92\%$ \\\cline{2-7}
                                        & \textit{8} & $474$ & $430$ & $9.28\%$ & $342$ & $27.85\%$ \\\Xhline{4\arrayrulewidth}
                                        
        \multirow{5}{*}{Four-Body SYK} & \textit{3} & $140$ & $60$ & $57.14\%$ & $60$ & $57.14\%$ \\\cline{2-7}
                                            & \textit{4} & $496$ & $432$ & $12.90\%$ & $312$ & $37.10\%$ \\\cline{2-7}
                                            & \textit{5} & $1848$ & $1440$ & $22.08\%$ & $896$ & $51.52\%$ \\\cline{2-7}
                                            & \textit{6} & $4760$ & $2568$ & $46.05\%$ & $2440$ & $48.74\%$ \\\cline{2-7}
                                            & \textit{7} & $9876$ & $6148$ & $37.75\%$ & $4988$ & $49.50\%$ \\\hline
         
    \end{tabular}
    \label{tab:hamiltonian-pauli-weight-exp}
\end{table*}

\begin{figure}[b]
    \centering
    \includegraphics[width=\linewidth]{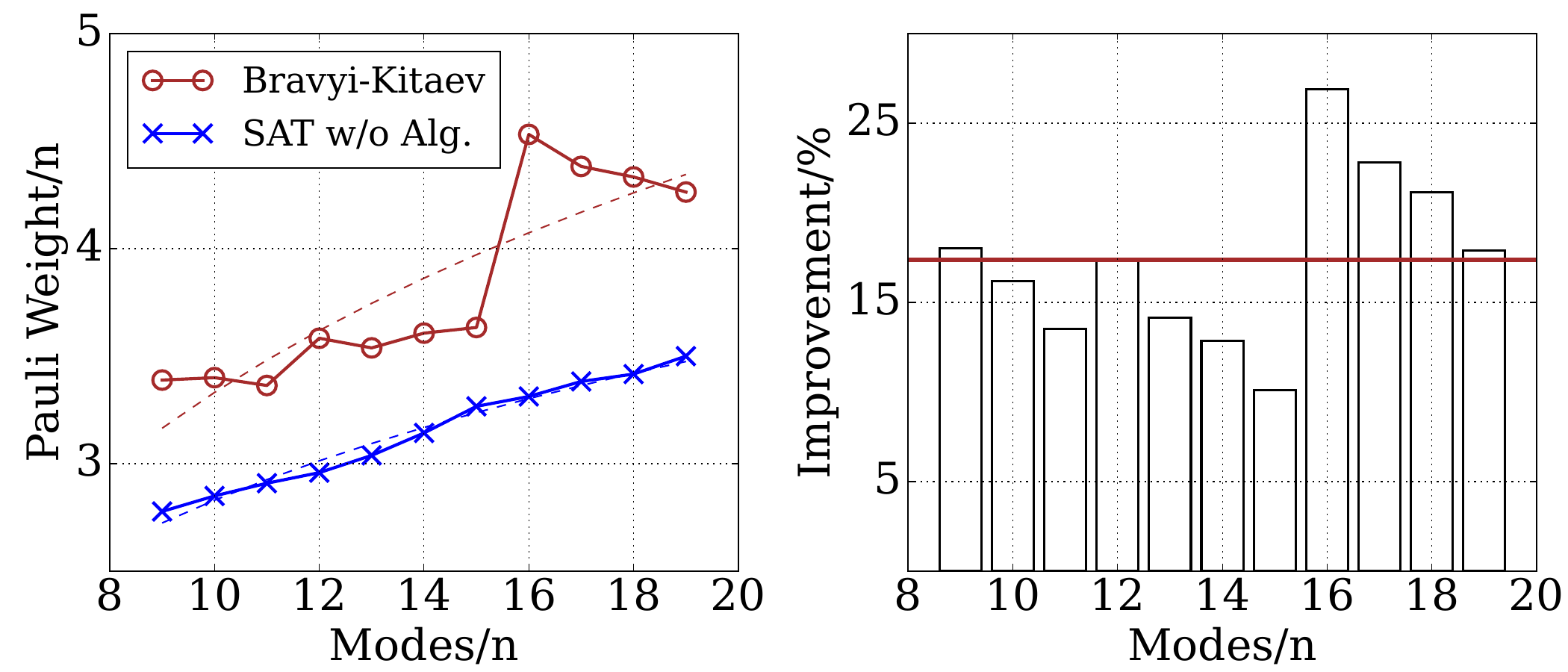}
    \caption{Average Pauli weight per-Majorana operator (larger scale)}
    \label{fig:plain-pauli-weight}
\end{figure}

We then evaluate the larger scale cases (9 to 19 Fermionic modes) where we compare \textbf{\textsf{SAT w/o Alg.}} with BK (\textbf{\textsf{Full SAT}} requires too many clauses for algebraic independence at this range). The average Pauli weight per Majorana operator is shown in Figure~\ref{fig:plain-pauli-weight}.
It can be observed that both BK and our method demonstrate a $O(\log N)$ per-Majorana operator Pauli weight, while our method is consistently smaller. The performance of BK varies for different numbers of Fermions, while our method can deliver the optimal performance stably. On average, our method can reduce the Pauli weight by $\mathbf{17.36\%}$ at 9-19 Fermionic modes while our regression coefficient indicates that we can expect around $\mathbf{36.38\%}$ reduction in per-Majorana operator Pauli weight between BK and the optimal Pauli weight as the number of Fermions increases.



\subsection{Hamiltonian-Dependent Encoding}

We then evaluate the Pauli weight for Hamiltonian-dependent Fermion-to-qubit encoding. 
We still start from small-size cases where \textbf{\textsf{Full SAT}} can be applied.
Table~\ref{tab:hamiltonian-pauli-weight-exp} shows the total Pauli weight of the three types of Hamiltonian benchmarks after using the BK encoding and our SAT-based encoding.
The \textbf{\textsf{Full SAT}} can consistently outperform BK with an average $37.26\%$ Pauli weight reduction.
The \textbf{\textsf{SAT+Anl.}} is suboptimal and can provide $21.63\%$ Pauli weight reduction on average against BK.
Notice that the \textbf{\textsf{SAT+Anl.}} is only worse than BK on extremely small cases with merely four Fermionic modes. 
This does not matter because we can always use  \textbf{\textsf{Full SAT}} as this scale.
For large-size cases, \textbf{\textsf{SAT+Anl.}} is constantly better than BK.
For those cases of 9 to 19 Fermionic modes, we compare BK with \textbf{\textsf{SAT+Anl.}}. The results are shown in Table~\ref{tab:hamiltonian-pauli-weight-annealing}. Compared with BK, \textbf{\textsf{SAT+Anl.}} can reduce the total Pauli weight by $23.71\%$ on average (up to $40\%$).


\begin{table}[t]
    \centering
    \caption{Hamiltonian Pauli weight of different problems (continue, \textbf{\textsf{SAT+Anl.}} only)}
    \begin{tabular}{|c|c|c|c|c|}\hline
        Case & $N$ & BK & \textbf{\textsf{SAT+Anl.}} & Reduction \\\hline
        \multirow{3}{*}{\makecell{Electronic\\Structure}} & \textit{8} & $24310$ & $22210$ & $8.64\%$ \\\cline{2-5}
         & \textit{10} & $54156$ & $43618$ & $19.46\%$ \\\cline{2-5}
         & \textit{12} & $151548$ & $117534$ & $22.44\%$ \\\Xhline{4\arrayrulewidth}
         
         \multirow{5}{*}{\makecell{Fermi\\-Hubbard}} & \textit{10} & $876$ & $682$ & $22.15\%$ \\\cline{2-5}
                                        & \textit{12} & $1404$ & $1094$ & $22.08\%$ \\\cline{2-5}
                                        & \textit{14} & $1978$ & $1488$ & $24.77\%$ \\\cline{2-5}
                                        & \textit{16} & $2626$ & $2216$ & $15.61\%$ \\\cline{2-5}
                                        & \textit{18} & $3436$ & $2738$ & $20.31\%$ \\\Xhline{4\arrayrulewidth}
                                        
         \multirow{4}{*}{\makecell{Four-Body \\ SYK}} & \textit{8} & $17376$ & $12848$ & $26.06\%$ \\\cline{2-5}
         & \textit{9} & $31952$ & $23328$ & $26.99\%$ \\\cline{2-5}
         & \textit{10} & $55208$ & $32976$ & $40.27\%$ \\\cline{2-5}
         & \textit{11} & $85436$ & $54924$ & $35.71\%$ \\\hline
    \end{tabular}
    \label{tab:hamiltonian-pauli-weight-annealing}
\end{table}

\subsection{End-to-End Real Hamiltonian Simulation}
To understand the end-to-end benefit of optimal Fermion-to-qubit encoding, 
we evaluate executing quantum Hamiltonian simulation programs with noisy simulation and a real quantum computer.
We select three benchmarks: the $H_2$ molecules (4 qubits), the $3\times1$ (6 qubits), and $2\times 2$ (8 qubits) Fermi-Hubbard Models with periodic boundary conditions. We compare three encodings: Jordan-Wigner (JW~\cite{jordan_uber_1928}), Bravyi-Kitaev (BK~\cite{bravyi_fermionic_2002}), and our \textbf{\textsf{Full SAT}}. The initial state is prepared to the energy eigenstate. 
All the generated Hamiltonians are optimized and compiled into quantum circuits with Paulihedral~\cite{li2022paulihedral} and Qiskit level 3 optimization~\cite{qiskit} with evolution time $t=1$.
For the real system study, we only evaluated $H_2$ molecule on the IonQ Aria-1 quantum computer due to the limited fidelity of available real quantum computers.

\subsubsection{Compilation Output}

Table~\ref{tab:gates-compiled} shows the gate count in the final compiled quantum circuits from different Fermion-to-qubit encoding methods, where \textit{Single} refers to the single-qubit gate and \textit{CNOT} is the two-qubit CNOT gate. On average, our \textbf{\textsf{Full SAT}} shows around $\sim20\%$ reduction on single qubit gates and $\sim35\%$ reduction on CNOT gates compared with BK. The significant reduction will later translate to an increase in simulation accuracy.
\begin{table}[t]
    \centering
    \caption{Gate count of compiled quantum circuits}
    \begin{tabular}{|c|c|c|c|c|}\hline
        Case & Gates & BK & \textbf{\textsf{Full SAT}} & Reduction \\\hline
        \multirow{4}{*}{$H_2$} & \textit{Single} & 26 & 22 & $7.69\%$ \\\cline{2-5}
        & \textit{CNOT} & 26 & 21 & $19.23\%$ \\\cline{2-5}
        & Total & 52 & 43 & $17.31\%$ \\\cline{2-5}
        & Depth & 39 & 33 & $15.38\%$ \\\Xhline{4\arrayrulewidth}

        \multirow{4}{*}{\makecell{$3\times1$\\Fermi-\\Hubbard}} & \textit{Single} & 58 & 41 & $29.31\%$ \\\cline{2-5}
        & \textit{CNOT} & 56 & 31 & $44.64\%$ \\\cline{2-5}
        & Total & 114 & 72 & $36.84\%$ \\\cline{2-5}
        & Depth & 76  & 26 & $65.79\%$ \\\Xhline{4\arrayrulewidth}
        
        \multirow{4}{*}{\makecell{$2\times2$\\Fermi-\\Hubbard}} & \textit{Single} & 56 & 41 & $26.79\%$ \\\cline{2-5}
        & \textit{CNOT} & 53 & 31 & $41.51\%$ \\\cline{2-5}
        & Total & 109 & 72 & $33.94\%$ \\\cline{2-5}
        & Depth & 73  & 26 & $64.38\%$ \\\hline
    \end{tabular}
    \label{tab:gates-compiled}
\end{table}



\subsubsection{Noisy Simulation}

We simulated the time evolution of $H_2$ and both Fermi-Hubbard models from different energy eigenstates under noise, where single-qubit gate fidelity is fixed to $99.99\%$, and double-qubits gate varies from $99.99\%$ to $99\%$. We re-run the circuit for multiple shots for each fidelity setting for a more precise calculation of the observed energy and its variance. The number of shots is $3000$ for $H_2$ and $1000$ for Fermi-Hubbard under each setting. 

Our noisy simulation starts from the energy eigenstates. Energy eigenstates are stationary under time evolution, meaning the system should retain and measure the same energy after a while. However, due to noise in the quantum circuit, the system would inevitably shift to other states for some small probabilities and cause the observed energy to drift. Lower noise should reduce such drifting and variance in measuring the final energy.

\begin{figure}[t]
    \centering
    \includegraphics[width=\linewidth]{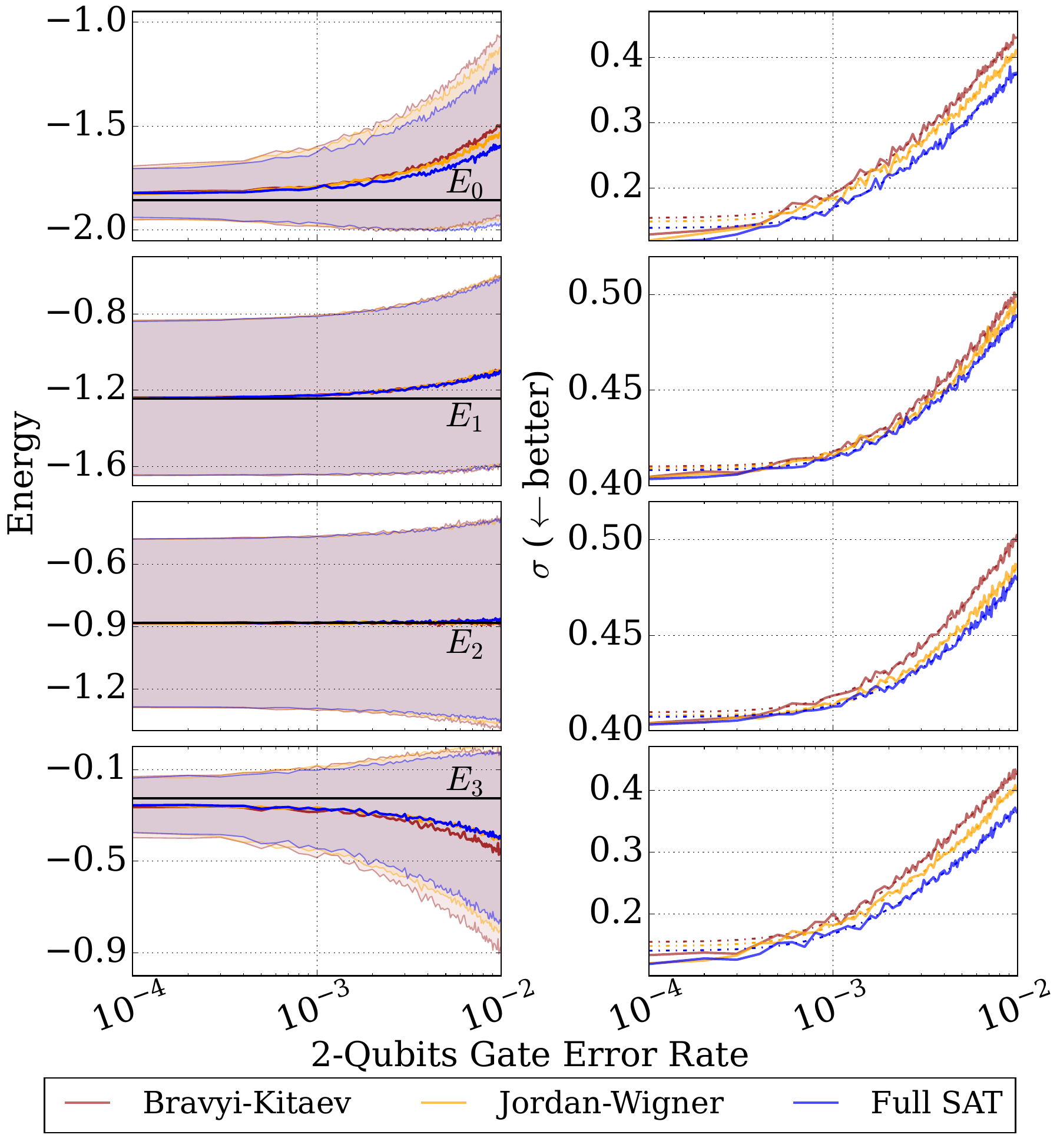}
    \caption{Noisy simulation of $H_2$, initial state $E_0\sim E_3$ (black line marks the theoretical energy, shadow marks $\pm1\sigma$ measurement S.D.)}
    \label{fig:noisy-simulation-H2}
\end{figure}

\textbf{$\mathbf{H_2}$}: Figure~\ref{fig:noisy-simulation-H2} shows the simulation result of $H_2$  molecule. Each row shows the result starting from a different energy eigenstate ($E_0$ to $E_3$). 
The X-axis is the 2-qubit gate error.
The Y-axis is the measured energy after the time evolution circuit.
The shadows in the left column represent the standard deviation of the measured energy, which is also plotted in the right column.
It can be observed that \textbf{\textsf{Full SAT}} can outperform BK and JW with the lowest drifting (closer to the energy level) and measuring variance thanks to the fewer gate counts due to lower Pauli weight in the optimal encoding.

\begin{figure}[t]
    \centering
    \includegraphics[width=\linewidth]{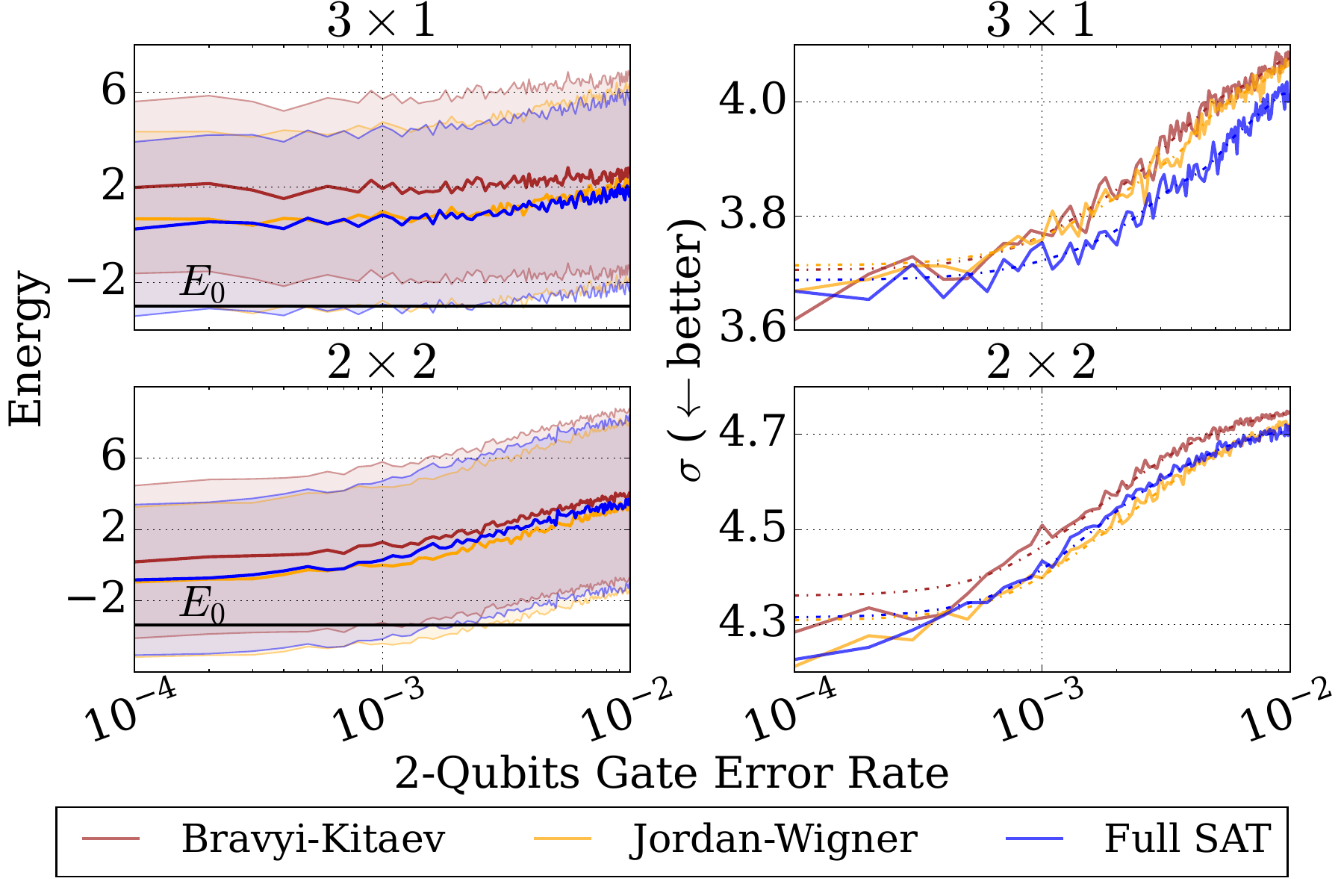}
    \caption{Noisy simulation of $3\times1$ and $2\times2$ Fermi-Hubbard Model, initial state $E_0$ (black line marks the theoretical energy, shadow marks $\pm1\sigma$ measurement S.D.)}
    \label{fig:noisy-simulation-fermi-hubbard}
\end{figure}

\textbf{Fermi-Hubbard}: Figure~\ref{fig:noisy-simulation-fermi-hubbard} shows the simulation result of $3\times 1$ and $2\times 2$ square lattice Fermi-Hubbard model with periodic boundary condition. 
For both models, we start the simulation from the ground energy eigenstate.
The representation of the X/Y-axis is the same as $H_2$.
Similarly, \textbf{\textsf{Full SAT}} can demonstrate the lower drifting. 

\subsubsection{Real System Study}
\begin{figure}[t]
    \centering
    \includegraphics[width=0.8\linewidth]{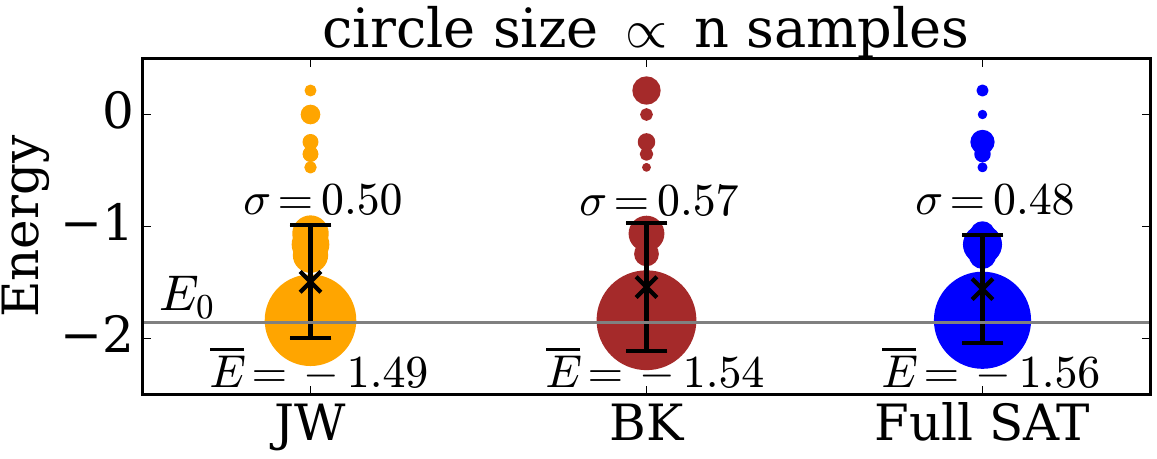}
    \caption{Measured energy of $H_2$ simulation on IonQ Aria-1, initial state $E_0$ (gray line)}
    \label{fig:ionq-H2}
\end{figure}

Figure~\ref{fig:ionq-H2} shows the time evolution of $H_2$ molecule from ground energy $E_0$ eigenstate on IonQ Aria-1 quantum computer.
The Y-axis is the measured energy. The circle size represents the number sampled to a certain value.
\textbf{\textsf{Full SAT}} can achieve the closest average energy ($-1.56$ vs. $-1.54$ vs. $-1.49$) with the smallest variance compared with BK and JW.

\subsection{Time-to-Solution and Scalability Discussion}

\begin{figure}[t]
    \centering
    \includegraphics[width=0.95\linewidth]{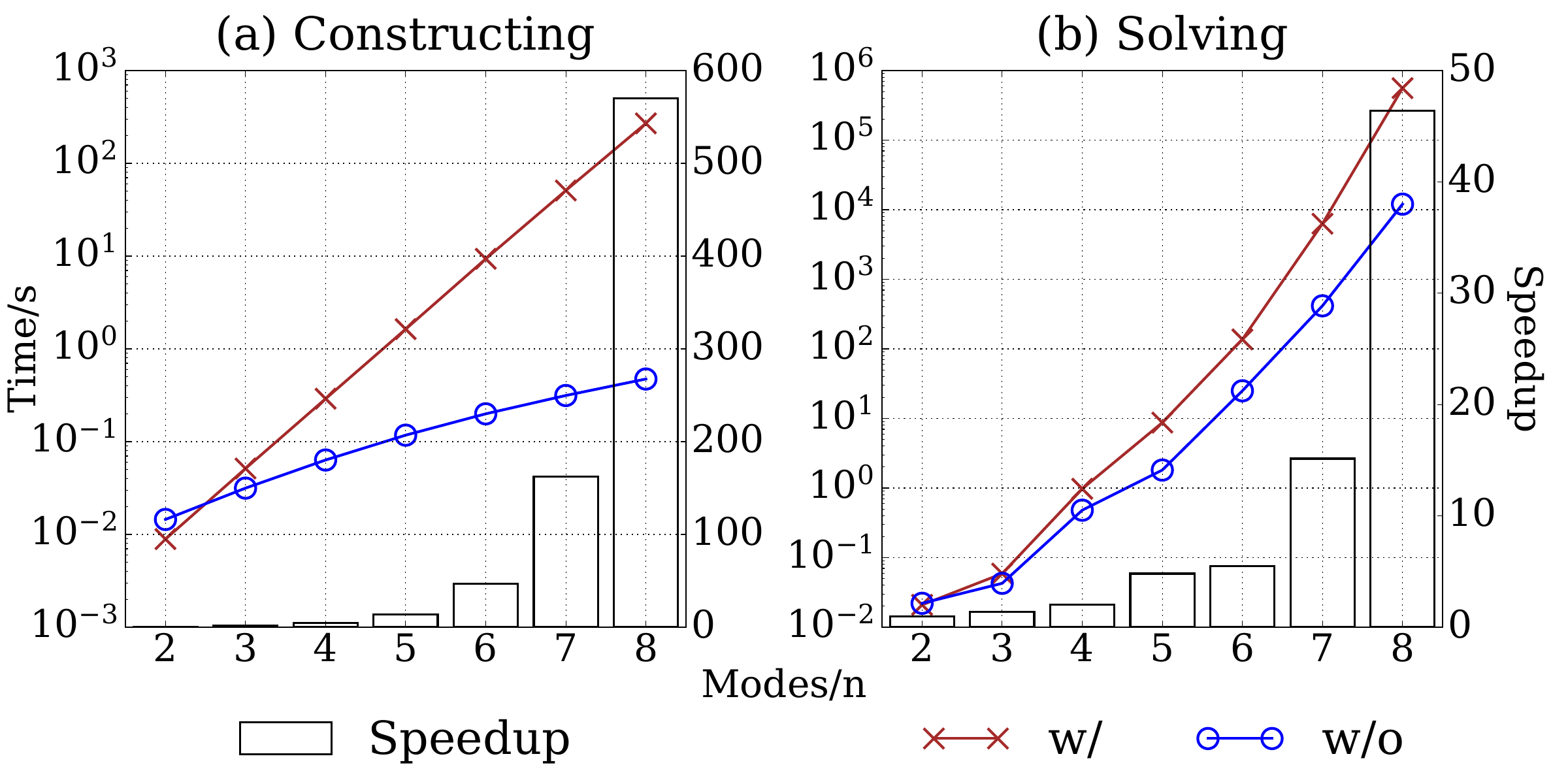}
    \caption{Time to construct and solve \textsf{\textbf{w, w/o Alg.}} (Proving unsatisfactory time is excluded)}
    \label{fig:benchmark}
\end{figure}

We also compare the time cost to solve the problem with and without algebraic independence using the SAT solver.
Figure~\ref{fig:benchmark} (a), (b) shows the time consumption of constructing and solving the SAT problem with/without algebraic independence.
Note that we exclude the time for the SAT solver to prove that a Pauli weight lower than optimal is unsatisfiable since it usually triggers a fixed timeout termination. Removing the algebraic independence constraints significantly reduced the time to construct and solve the SAT problem.

Scalability is a major concern of SAT as it is NP-complete, and the time-to-solution, even on small scales, is rather long. However, we only apply a single-thread SAT solver on a small server. Employing a distributed SAT solver on a high-performance cluster could help solve larger-size cases, but confirming such speed-up requires further experiments. More importantly, our method can reveal the structure of optimal Fermion-to-qubit encoding for different Hamiltonians and thus guide future work in similar directions, e.g., approximate-optimal encodings with lower complexity, archi-tecture-aware compilations, etc.

\section{Related Works}

\textbf{Fermion-to-Qubit Encoding}: 
Previously, Fermion-to-qubit encoding has been studied from a theoretical and constructive perspective. General encoding schemes include Jordan-Wigner transformation \cite{jordan_uber_1928}, Bravyi-Kitaev transformation~\cite{bravyi_fermionic_2002}, Parity~\cite{bravyi2017tapering-parity}, ternary tree~\cite{Jiang_2020,miller_bonsai_2022}, and they have been widely implemented in many quantum software frameworks like Qiskit~\cite{qiskit} and OpenFermion~\cite{openfermion}.
Among them, the Bravyi-Kitaev transformation~\cite{bravyi_fermionic_2002} (\cite{bk-compare-with-jw} compares Bravyi-Kitaev with Jordan-Wigner) and ternary tree~\cite{Jiang_2020,miller_bonsai_2022} have achieved the asymptotical optimal Pauli weight per Majorana operator $O(\log N)$ for $N$ Fermionic modes.
But they are still far from the actual optimal solutions.
Moreover, the actual Hamiltonian structure is not considered in these approaches.
This paper formulates the entire Fermion-to-qubit encoding along with the Hamiltonian-dependent cost into an SAT problem and thus can provide the optimal solution. Recently, the superfast encoding~\cite{superfast, superfast-Chien_2019} is proposed for a particular type of Fermionic systems with only local Fermionic mode and cyclic structured interactions. This work, however, can process any weakly (Fermi-Hubbard) and strongly (SYK, electronic structure) interacted systems, especially since it considers the Hamiltonian structure in the SAT formulation.

\textbf{Generic Quantum Compiler Optimization}: 
Today's quantum compilers (e.g., Qiskit~\cite{qiskit}, Cirq~\cite{cirq}, Q\#~\cite{qsharp}) usually run multiple passes to optimize a quantum circuit. Typical passes include gate cancellation~\cite{gatecancel}, gate rewrite~\cite{gaterewrite}, and routing~\cite{routing_sat, murali2019noiseadaptive}. These optimization methods (passes) are usually small-scale local circuit transformations. They cannot optimize the Fermion-to-qubit encoding as it will rewrite the entire Hamiltonian simulation circuit.

\textbf{Compilation for Quantum Simulation}:
Recently, several works have identified domain-specific compiler optimization opportunities in quantum simulation.
These optimizations include the Pauli string ordering~\cite{sim-opt-gui2021term,sim-opt-hastings}, architec-tural-aware synthesis for Pauli string~\cite{sim-opt-degriend2020architectureaware, li2022paulihedral}, simultaneous diagonalization on commutative Pauli strings~\cite{diag_van_den_Berg_2020, diag_Cowtan_2020, diag_cowtan2020generic}. All these works happen at the Pauli string level after finishing the Fermion-to-qubit encoding. This work optimizes the Fermion-to-qubit encoding, which happens before the simulation problem is turned into Pauli strings, and can naturally be combined with all these works.

\textbf{Formal Methods in Quantum Compilation}: Several previous works bring the SAT method to qubit mapping and routing on connectivity-constrained architectures by different constraint encoding, including \cite{routing_sat, Tan_2021, Tan_2020,robert2009,10.1145/3316781.3317859}, and \cite{Murali_2019} (SMT based). 
These methods focus on different aspects (fidelity, gate count, and circuit depth) and still optimize the gates in the final quantum circuit.
To the best of our knowledge, this work is the first to formalize the Fermion-to-qubit encoding finding into an SAT problem.


\section{Conclusion}

In summary, this paper introduces \myFrameworkName, a novel compiler framework that addresses the challenge of efficiently simulating complex Fermionic systems on quantum computers. 
\myFrameworkName converts the problem of finding optimal Fermion-to-qubit encoding into an SAT problem with carefully designed constraint/objective encoding and solves it with high-performance SAT solvers.
Moreover, it offers approximate optimal outcomes for larger-scale scenarios by overcoming the complexities of the exponentially large number of clauses involved in this encoding process. The extensive evaluation across diverse Fermionic systems shows its superiority over existing methods, notably reducing implementation costs and enhancing efficiency in quantum simulations.
This work marks a substantial step forward, providing a robust solution for encoding Fermionic systems onto quantum computers and advancing quantum simulation capabilities across scientific domains.

\begin{acks}
We thank the anonymous reviewers for their insightful feedback and our shepherd Yipeng Huang for guidance. This work was in part supported by NSF CAREER Award 2338773 and Amazon Web Service Cloud Credit.
\end{acks}

\appendix
\section{Artifact Appendix}

\subsection{Abstract}

The artifact contains the code of our Fermihedral framework and other scripts to prepare the environment and reproduce our key results and figures (Figures~\ref{fig:probability-hold},~\ref{fig:small-scale-average},~\ref{fig:plain-pauli-weight},~\ref{fig:noisy-simulation-H2},~\ref{fig:noisy-simulation-fermi-hubbard}, and~\ref{fig:ionq-H2}; Tables~\ref{tab:hamiltonian-pauli-weight-exp} and~\ref{tab:hamiltonian-pauli-weight-annealing}. It requires a regular x86-64 Linux server with at least 32GB RAM and 10GB available harddrive. A CUDA GPU is recommended to speed up some of the experiments. Gate counts in Table 5 cannot be automatically reproduced due to break changes and version conflict in Qiskit with Paulihedral. We also provide instructions on how to run the circuit on the IonQ quantum computer device, which requires extra manual work, as well as the entire software dependencies list and generated Hamiltonian model. Since randomization is applied in simulated annealing, the result in Tables 3 and 4 could be deviated.

\subsection{Artifact check-list (meta-information)}

{\small
\begin{itemize}
  \item {\bf Algorithm: } Fermihedral has following core algorithms: \begin{itemize}
      \item Solving the Fermion-to-qubit encoding problem with SAT (Section~\ref{sec:main-SAT-method}) is implemented in the file `fermihedral/majorana.py', \verb|DescentSolver| and \verb|MajoranaModel|.
      \item Solving Hamiltonian-dependent Pauli weight with \textbf{\textsf{Full SAT}} or \textbf{\textsf{SAT+Anl.}} method (Section~\ref{sec:hamiltonian-pauli-weight}) is implemented in the file `fermihedral/majorana.py', \verb|HamiltonianSolver|.
  \end{itemize}
  \item {\bf Program: } Noisy simulation subroutines are implemented in the file `fermihedral/fock.py`.
  \item {\bf Run-time environment: } Python, Jupyter Notebook
  \item {\bf Hardware: } Single x86-64 CPU Linux server, preferably with a CUDA GPU, to run the noisy simulation faster.
  \item {\bf Metrics: } We consider the following metrics and could be directly observed from the output figures: \begin{itemize}
      \item Pauli weight
      \item Observed system energy
  \end{itemize}
  \item {\bf Output: } The output contains the Fermion-to-qubit encoding schema. The output is adapted to the \verb|FermionicMapper| interface in Qiskit.
  \item {\bf Experiments: } There are three notebooks for different sets of experiments: \begin{enumerate}
      \item \verb|singleshot.ipynb|: Figures~\ref{fig:probability-hold},~\ref{fig:small-scale-average}, and~\ref{fig:plain-pauli-weight}. The time to run this notebook could be extremely long.
      \item \verb|simulation.ipynb|: Figures~\ref{fig:noisy-simulation-H2},~\ref{fig:noisy-simulation-fermi-hubbard}, and~\ref{fig:ionq-H2}.
      \item \verb|hamiltonian-weight.ipynb|: Table~\ref{tab:hamiltonian-pauli-weight-exp} and~\ref{tab:hamiltonian-pauli-weight-annealing}. The time to run this notebook could be extremely long.
  \end{enumerate}
  \item {\bf Disk space required: } 10GB
  \item {\bf Time to prepare workflow: } 5 minutes
  \item {\bf Time to complete experiments: } 1 to 2 weeks
  \item {\bf Publicly available: } Yes
  \item {\bf Code licenses: } MIT License
  \item {\bf Workflow framework used: } Python, Qiskit, Jupyter Notebook
  \item {\bf Archived: } 10.5281/zenodo.10854557
\end{itemize}
}

\subsection{Description}

\subsubsection{How to access}

The artifact is available at the following GitHub repository \url{https://github.com/acasta-yhliu/fermihedral.git} or with DOI \url{https://doi.org/10.5281/zenodo.10854557}. You can clone the repository or submit issues if there is any problem.

\subsubsection{Hardware dependencies}

A regular server with a single CPU and preferably a CUDA GPU can run our artifact. RAM is not strictly limited but is recommended to 32GB or above. We also recommend using a powerful CPU to speed up the SAT solver. 

A small part of our results (Figure~\ref{fig:ionq-H2}) requires execution on the real IonQ quantum computer (available through Amazon Web Service). If you wish to run the quantum circuit simulating $H_2$ molecule on this real IonQ quantum computer, the cost should be around \$600 (charged by Amazon) with our setup. Consequently, we exclude this part from artifact evaluation, and the exclusion will not affect the major results of this paper.

\subsubsection{Software dependencies}

The artifact is implement-ed in Python 3.10. We also require Python packages, including z3-solver 4.12.2.0, Qiskit 1.0.1 accompanied by Qiskit-nature 0.7.2, Qiskit-algorithm 0.3.0, and OpenFermion 1.5.1. Both Qiskit packages require Numpy 1.25.2. The list of Python packages and other assistant packages can be found in the \verb|requirements.txt|.

We require a SAT solver to solve the model. To install and compile the \verb|kissat| SAT solver, \verb|make| and a C compiler are required. We used GCC 11.4.0 in our experiments.

\subsection{Installation}

You can clone the repository to your local machine and prepare the environment needed by our artifact with the following command (script):
\begin{verbatim}
$ python3 prepare.py
\end{verbatim}

\subsection{Experiment workflow}

After cloning the repository and preparing the environment in the installation setup, you can execute the Jupyter Notebooks in the virtual environment to reproduce the result (order does not matter).
\begin{enumerate}
  \item \verb|singleshot.ipynb|: Figures~\ref{fig:probability-hold},~\ref{fig:small-scale-average}, and~\ref{fig:plain-pauli-weight}.
  \item \verb|simulation.ipynb|: Figures~\ref{fig:noisy-simulation-H2},~\ref{fig:noisy-simulation-fermi-hubbard}, and~\ref{fig:ionq-H2}.
  \item \verb|hamiltonian-weight.ipynb|: Table~\ref{tab:hamiltonian-pauli-weight-exp} and~\ref{tab:hamiltonian-pauli-weight-annealing}.
\end{enumerate}
The results are produced in the exact order of the figures and tables. Please note that the time to execute experiments in \verb|singleshot.ipynb| and \verb|hamiltonian-weight.ipynb| could be extremely long (around weeks).

\subsection{Evaluation and expected results}

Notebooks should reproduce the exact figures as in our paper. The results are directly printed for Tables~\ref{tab:hamiltonian-pauli-weight-exp} and~\ref{tab:hamiltonian-pauli-weight-annealing}.

\subsection{Experiment customization}

If a CUDA GPU is available, uncomment the following line in the first code cell in \verb|simulation.ipynb| to enable GPU acceleration:
\begin{verbatim}
# uncomment this line to enable GPU support
# config_device("GPU")
\end{verbatim}

Solving the SAT problem at a large scale takes an extremely long time. If the time is a problem, it is possible to limit the problem scale in \verb|singleshot.ipynb| by lowering the number of \verb|MAX_MODES| at the beginning of cells. The timeout number of the SAT solver can also be adjusted to fit the local machine's performance.

For \verb|hamiltonian-weight.ipynb|, you can also adjust the models it solves. \verb|FULL_SAT_MODELS| controls \textbf{\textsf{Full SAT}} method and \verb|ANNEALING_MODELS| controls \textbf{\textsf{SAT+Anl.}} method. Your model must follow the format:
\begin{enumerate}
    \item First line marks the model name, modes, and format (\verb|ac| or \verb|mj|). Hamiltonians characterized by creation and annihilation operators are marked by \verb|ac|, while \verb|mj| indicates Majorana operators characterize the Hamiltonian.
    \item Each line is considered as a production of operators. Creation operators are positive numbers, while annihilation operators are negative numbers. Majorana operators are all positive operators.
\end{enumerate}

\bibliographystyle{plain}
\bibliography{references}

\end{document}